\documentclass[preprint2]{aastex}

\newcommand{\bootes}{Bo\"otes}

\slugcomment{Preprint to be submitted to ApJ}

\shorttitle{The Infrared Luminosity Function of Quasars}
\shortauthors{Brown {\it et al.}}

\begin{document}

\title{The {\boldmath $1<z<5$} Infrared Luminosity Function of Type I Quasars}

\author{
Michael J. I. Brown\altaffilmark{1\dagger}, 
Kate Brand\altaffilmark{2}, 
Arjun Dey\altaffilmark{2}, 
Buell T. Jannuzi\altaffilmark{2}, 
Richard Cool\altaffilmark{3},
Emeric Le~Floc'h\altaffilmark{3,4},
Christopher S. Kochanek\altaffilmark{5},
Lee Armus\altaffilmark{6},
Chao Bian\altaffilmark{7},
Jim Higdon\altaffilmark{8},
Sarah Higdon\altaffilmark{8},
Casey Papovich\altaffilmark{3},
George Rieke\altaffilmark{3},
Marcia Rieke\altaffilmark{3},
J. D. Smith\altaffilmark{3},
B. T. Soifer\altaffilmark{6,7},
Dan Weedman\altaffilmark{8}}

\altaffiltext{1}{Princeton University Observatory, Peyton Hall, Princeton, NJ 08544-1001, USA}
\altaffiltext{2}{National Optical Astronomy Observatory, Tucson, AZ 85726-6732, USA}
\altaffiltext{3}{Steward Observatory, University of Arizona, Tucson, AZ 85721, USA}
\altaffiltext{4}{Observatoire de Paris, GEPI, 92195 Meudon, France}
\altaffiltext{5}{Department of Astronomy, The Ohio State University, 140 West 18th Avenue, Columbus, OH 43210, USA}
\altaffiltext{6}{Spitzer Science Center, IPAC, California Institute of Technology, MC 220-6, 1200 East California Boulevard, Pasadena, CA 91125, USA}
\altaffiltext{7}{The Division of Physics, Mathematics \& Astronomy, California Institute of Technology, 1200 East California Boulevard, Pasadena, CA 91125, USA}
\altaffiltext{8}{Astronomy Department, Cornell University, Ithaca, NY 14853, USA}
\altaffiltext{$\dagger$}{H.N. Russell Fellow\vspace{0.75cm}}

\email{mbrown@astro.princeton.edu}

\begin{abstract}
We determine the rest-frame $8\micron$ luminosity function of type I quasars over
the redshift range $1 < z < 5$.  Our sample consists of 292 $24\micron$ sources             
brighter than $1$~mJy selected from $7.17~{\rm deg^2}$ of the {\it Spitzer Space Telescope} MIPS survey        
of the NOAO Deep Wide-Field Survey \bootes~ field.  The AGN and Galaxy Evolution
Survey (AGES) has measured redshifts for 270 of the $R<21.7$ sources and we estimate that
the contamination of the remaining 22 sources by stars and galaxies is low.                 
We are able to select quasars missed by ultra-violet excess quasar surveys,
including reddened type I quasars and $2.2<z<3.0$ quasars with optical colors similar
to main sequence stars. We find reddened type I quasars comprise 
$\sim 20\%$ of the type I quasar population. Nonetheless, the shape, normalization, 
and evolution of the rest-frame $8\micron$ luminosity function is comparable 
to that of quasars selected from optical surveys. The $8~\micron$ luminosity function of 
type I quasars is well approximated by a power-law with index $-2.75\pm 0.14$.
We directly measure the peak of the quasar space density to be at $z=2.6\pm 0.3$.           
\end{abstract}


\section{INTRODUCTION}
\label{sec:intro}

The luminosity function of Active Galactic Nuclei (AGNs) is 
one of the principal constraints on the growth of supermassive 
black holes over the age of the Universe \markcite{kau00}(e.g., {Kauffmann} \& {Haehnelt} 2000). 
Ideally, we would measure the bolometric luminosity function of 
all AGNs as a function of redshift. This is not trivial, as it requires deep 
multiwavelength data spanning X-ray to sub-mm wavelengths and 
sampling large comoving volumes.  In practice, surveys
select subsets of AGNs in relatively narrow wavelength ranges. 
For example, optical quasar surveys generally select quasars
with blue spectral energy distributions and broad emission lines.
While using a narrow wavelength range as a proxy for
bolometric luminosity would appear risky, this approach
seems to have been remarkably successful. Optical, radio,
and X-ray luminosity functions of AGNs infer
a $z\simeq 2.5$ peak in the space density of the most luminous quasars
\markcite{dun90,ued03,cro04}(e.g., {Dunlop} \& {Peacock} 1990; {Ueda} {et~al.} 2003; {Croom} {et~al.} 2004). This agreement is reassuring,
since the three selection methods have different biases --
strong radio emission only occurs in a small subset of all AGNs,
optical wavelengths are sensitive to dust absorption, and soft X-ray wavelengths
are sensitive to gas absorption.

It is the absorption of radiation by the gas and dust surrounding
the black hole that creates the greatest systematic uncertainties.
In particular, unified models of AGNs \markcite{ant93}(e.g., {Antonucci} 1993) assume
the presence of obscuring material with a non-isotropic spatial 
distribution that makes the observed spectrum
a function of viewing angle. In unified models of AGNs, 
the obscuring material is often
referred to as the torus, though the actual distribution of gas and dust
may be significantly more complex and its properties could
depend on luminosity, time or Eddington factors \markcite{law91,hop05}(e.g., {Lawrence} 1991; {Hopkins} {et~al.} 2005).  
When AGNs are viewed from above the torus, optical continuum and
broad emission lines associated with regions near the 
central black hole can be observed directly. When AGNs 
are viewed in the plane of the torus, optical continuum
and the broad emission lines are absorbed. 

Selection biases due to absorption by dust or gas are minimized by
selecting AGNs in the infrared.  In this paper we combine 
{\it Spitzer Space Telescope} $24~\micron$ imaging 
of the  Bo\"otes field \markcite{rie04,hou05}({Rieke} {et~al.} 2004; {Houck} {et~al.} 2005) of the optical 
NOAO Deep Wide-Field Survey \markcite{jan99}(NDWFS; {Jannuzi} \& {Dey} 1999), with spectroscopy of quasar candidates
by the AGN and Galaxy Evolution Survey (AGES, Kochanek et al. in preparation)
to determine the rest-frame $8~\micron$ luminosity function of type I
quasars.  This allows us to identify quasars that would be
missed in optical surveys either because of extinction or
because the quasars have optical colors typical of main
sequence stars.  

The structure of this paper is as follows.
In \S\ref{sec:surveys} we describe the surveys used to 
produce the quasar sample.
In \S\ref{sec:t1} we characterize the sample,
discuss its completeness and compare it to other selection
methods available for the field.  In \S\ref{sec:lf} we derive the 
luminosity function of the type I quasars in the sample,
and in \S\ref{sec:obscured} we discuss the role of type II quasars.  
We summarize our results and discuss future prospects
in \S\ref{sec:sum}. Throughout this paper  
we define type I quasars as AGNs with optical and ultraviolet emission
lines broader than $1000~{\rm km}~{\rm s}^{-1}$ and type II quasars as those
with narrower lines. 
In this paper we adopt a cosmology of 
$\Omega_m=0.3$, $\Omega_\Lambda=0.7$, and $H_0=70~{\rm km}~{\rm s}^{-1}~{\rm Mpc}$.
Throughout this paper we use $AB$ magnitudes although 
the released NDWFS, 2MASS, and USNO-A2 catalogs use Vega 
magnitudes\footnote{$B_{W,AB}=B_{W,{\rm Vega}}$, 
$R_{AB}=R_{{\rm Vega}}+0.20$, 
$I_{AB}=I_{{\rm Vega}}+0.45$, 
and $J_{AB}=J_{{\rm Vega}}+0.89$}.

\section{THE SURVEYS}
\label{sec:surveys}

Our sample combines three surveys of \bootes~--~the 
NDWFS and {\it Spitzer} MIPS imaging surveys
and the AGES spectroscopic survey. In this section we discuss the
general properties of these three surveys and summarize 
the final multi-wavelength catalog. 

\subsection{The NOAO Deep Wide-Field Survey}
\label{sec:ndwfs}

The NDWFS is an optical and near-infrared imaging survey of two
$\approx 9.3~{\rm deg}^2$ fields with NOAO telescopes \markcite{jan99}({Jannuzi} \& {Dey} 1999). 
In this paper, we utilize the third data release of NDWFS $B_WRI$ imaging of the \bootes~ field,
which is available from the NOAO Science Archive\footnote{http://www.archive.noao.edu/ndwfs/}.
A thorough description of the optical observing strategy and data reduction
will be provided by Jannuzi et al. (in preparation).

Object catalogs were generated using SExtractor $2.3.2$ \markcite{ber96}({Bertin} \& {Arnouts} 1996)
run in single-image mode. Detections in the different bands 
were matched if the centroids were within  $1^{\prime\prime}$ of each other.
For extended objects, detections in the different bands were matched
if the centroid in one band was within an ellipse defined using the second
order moments of the light distribution of the object in another band\footnote{This
ellipse was defined with the SExtractor parameters $2 \times
{\rm A\_WORLD}$, $2 \times {\rm B\_WORLD}$, and ${\rm THETA\_WORLD}$.}.
Throughout this paper we use SExtractor MAG\_AUTO magnitudes
\markcite{kro80}(which are similar to Kron total magnitudes; {Kron} 1980) for optical photometry, 
due to their small uncertainties and systematic errors at faint magnitudes.
The 50\% completeness limits of the SExtractor catalogs for point sources are
$B_W\simeq 26.5$, $R\simeq 25.5$ and $I\simeq 25.3$. 

We used SExtractor's star-galaxy classifier to classify objects
as compact or extended. For AGES target selection, objects were
classified as compact if they had a CLASS\_STAR parameter of $\geq 0.8$ in 
one or more optical bands. This resulted in a small number
of emission line galaxies being erroneously classified as compact.
For this paper, we classify objects as compact if they have a
CLASS\_STAR of $\geq 0.7$ in two or more of the three optical bands.
Of the 288 $24~\micron$ sources with AGES spectroscopy which 
are classified as compact with our CLASS\_STAR criterion, 20 are 
$z<1$ galaxies which we reclassify as extended objects for the remainder of the paper.   
Type I quasar candidates are selected from the sample of $24~\micron$
sources with compact optical morphologies (\S\ref{sec:t1}). This reduces
galaxy contamination while maintaining a high completeness for 
$z>1$ quasars.  Of the 1182 $24~\micron$ sources with $R<20.2$                           
extended counterparts and AGES spectroscopy, only two are $z>1$ type I quasars.
We do not attempt to measure the $z<1$ type I quasar and Seyfert 
luminosity function, as the host galaxies of $z<1$ AGNs 
are sometimes detected and resolved in the deep NDWFS imaging resulting 
in a complex selection function that depends on host morphology, host luminosity,
AGN optical luminosity, redshift and image quality.
Our selection criteria also eliminates type II quasars from the sample,
which we will discuss in detail in a second paper. 
Our measurement of the 
$1<z<5$ type I quasar luminosity function and our conclusions are not
highly sensitive to our criteria for star-galaxy classification.

\subsection{The MIPS Survey}

The NDWFS \bootes~field was observed simultaneously at 24, 70, and $160~\micron$
with the Multiband Imaging Photometer for {\it Spitzer} \markcite{rie04}(MIPS; {Rieke} {et~al.} 2004) as part 
of the {\it Spitzer} IRS team's Guaranteed Time Observing programs \markcite{hou05}({Houck} {et~al.} 2005). 
In this paper we only discuss the $24~\micron$ data, as only five of the 
type I quasars in our sample were detected in the $70~\micron$ and $160~\micron$ imaging. 
These observations were performed with the ``Medium Scan'' technique. This observing mode is
particularly suitable for high efficiency coverage of wide fields
such as \bootes.  The $24~\micron$ detector uses a
$2\farcs55$ per pixel array of 128$\times$128 elements with 
a delivered point source full-width at half maximum (FWHM) of $\simeq 6^{\prime\prime}$.
The effective integration time per sky pixel was $\simeq 90~{\rm s}$.  Data
reduction was carried out using the MIPS Data Analysis Tool \markcite{gor05}({Gordon} {et~al.} 2005).
We corrected count rates for dark current, cosmic--rays, and flux non--linearities. 
These counts were then divided by flat--fields constructed for each position 
of the scan--mirror which is used in the ``Medium Scan" observing mode.  The images
were corrected for geometric distortions and combined to produce the final
$9.93~{\rm deg}^2$ $24~\micron$ mosaic.                                                    
For this work, we selected $24~\micron$ sources from the $8.22~{\rm deg^2}$ of the 
$24~\micron$ mosaic which overlaps the NDWFS and comprises of two or more individual scans.

As almost all the sources in the MIPS mosaic are unresolved, 
source extraction and photometry were performed using the stellar PSF fitting
technique of the DAOPHOT software \markcite{ste87}({Stetson} 1987). An empirical PSF 
was constructed from the brightest objects found
in our mosaic, and it was subsequently fitted to all the sources
detected in the map.  Allowing for multiple-match fitting to deal with
blended cases, we derived the flux density of each source from the
scaled fitted PSF and finally applied a slight correction to account
for the finite size of the modeled point spread function.
Catalog depth was determined by adding artificial point sources to the
MIPS mosaic and attempting to recover them with  DAOPHOT.
We estimate that the 80\% completeness limit to be $0.3~{\rm mJy}$.

\subsection{The AGN and Galaxy Evolution Survey}

The AGN and Galaxy Evolution Survey (AGES, Kochanek et al. in preparation) is obtaining
complete spectroscopic samples of galaxies and quasars selected over a broad range of
wavelengths in the Bo\"otes field of the NDWFS. The spectra are obtained
with Hectospec, a 300-fiber robotic spectrograph on the MMT 6.5-m
telescope \markcite{fab98,rol98}({Fabricant} {et~al.} 1998; {Roll}, {Fabricant}, \& {McLeod} 1998).   Spectra were extracted and classified
using two independent reduction pipelines and then verified by 
visual inspection.  One pipeline is based upon the SDSS spectroscopic
data reduction software while the other is a set of customized IRAF scripts.
The wavelength coverage of the spectra extends
from 3700\AA~to 9200\AA, although some spectra are contaminated 
redwards of 8500\AA~ by a red leak from an LED in the fiber positioner.

The AGES spectroscopic targets were selected from the April 9, 2004 
pre-release versions of the NDWFS and MIPS $24~\micron$ source catalogs. 
However, for the remainder of the paper we use the photometry of the third
data release of the NDWFS and July 5, 2005 pre-release version of the 
MIPS $24~\micron$ catalog. These have less artifacts and slightly better 
photometric calibrations than the earlier catalogs.
The spectra are mostly from AGES internal release 1.1 with some additional
spectra of $R$-band selected targets from AGES internal release 2.0. 
AGES targeted $24~\micron$ sources with fluxes of $S_{24}>1$~mJy 
(or $[24]<16.40$ where $[24]$ is the $AB$ apparent magnitude) which 
had optically extended $R<20.2$ counterparts or optically compact $17.2<R<21.7$
counterparts. For AGES target selection, objects were classified as 
compact if their SExtractor CLASS\_STAR parameter was $\geq 0.8$ in 
one or more of the $B_W$, $R$, or $I$-bands. To reduce the number of bright ($R\lesssim 18$)
galactic stars in the spectroscopic sample, compact sources were not 
targetted if they had 2MASS counterparts bluer than $J-[24]=-3.51$. 
The roughly power-law spectra of quasars and the black
body spectra of stars make the infrared color 
differences of these two populations very obvious \markcite{ste05}(e.g., {Stern} {et~al.} 2005). 
These criteria led to 1571 spectroscopic targets, and accurate 
spectroscopic redshifts were obtained for 1449 of them.                                   

\begin{figure}[hbt]
\epsscale{1.00}
\plotone{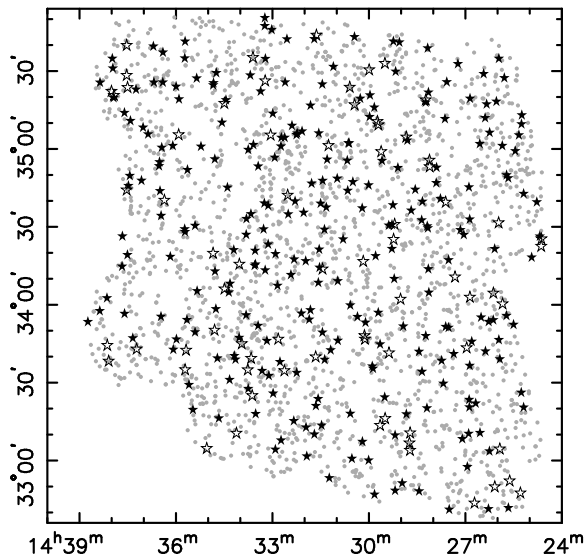}
\epsscale{1.00}
\caption{The sky distribution of $24~\micron$ sources brighter than $1~{\rm mJy}$ in the multiwavelength catalog.
Compact objects with and without spectra are shown with solid and open stars respectively.
Gray circles denote the locations of sources with extended optical morphologies.
The semicircular edges of the sample are an artifact of the circular 1~degree field
of view of the Hectospec instrument.
\label{fig:sky}}
\end{figure}

\subsection{The Multiwavelength Sample}

The final, multiwavelength catalog consists of $24~\micron$ sources 
brighter than $1~{\rm mJy}$ with good $24~\micron$ and optical 
imaging inside the primary AGES target regions.
For saturated objects in the NDWFS images, we have used the fourth 
data release of the Sloan Digital Sky Survey \markcite{ade05}(SDSS DR4; {Adelman-McCarthy} 2005) 
for the photometry and classifications. We excluded sources with 
$R>17.2$ counterparts within $(20.2-R)^{\prime\prime}$ of $R<15.2$ USNO-A2 
stars \markcite{mon98}({Monet} {et~al.} 1998) where our ability to detect faint optical counterparts is compromised.
We also excised $\gtrsim 1^{\prime}$ radius regions around Hipparcos stars 
\markcite{esa97}({Esa} 1997), which are occasionally broken up into multiple spurious
USNO-A2 catalog entries. This left us with an effective survey area of $7.17~{\rm deg^2}$.
Objects with spurious $R>21.7$ counterparts (e.g., edges of $R\sim 18$ 
star halos) have been removed from the final catalog.
The sky distribution of objects in the catalog is shown in Figure~\ref{fig:sky}. 

\begin{figure*}
\vspace{1cm}
\epsscale{0.90}
\plotone{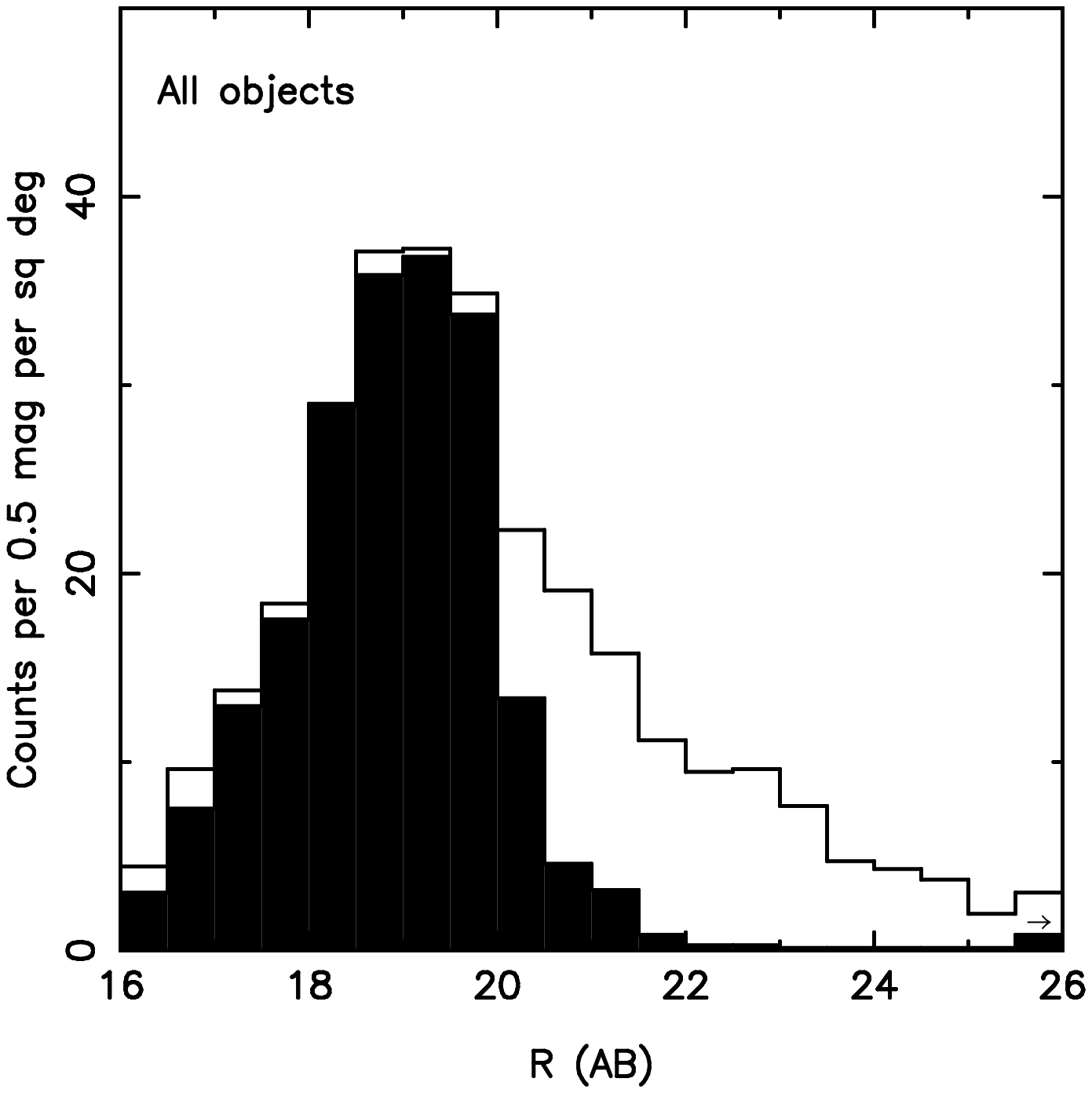}\plotone{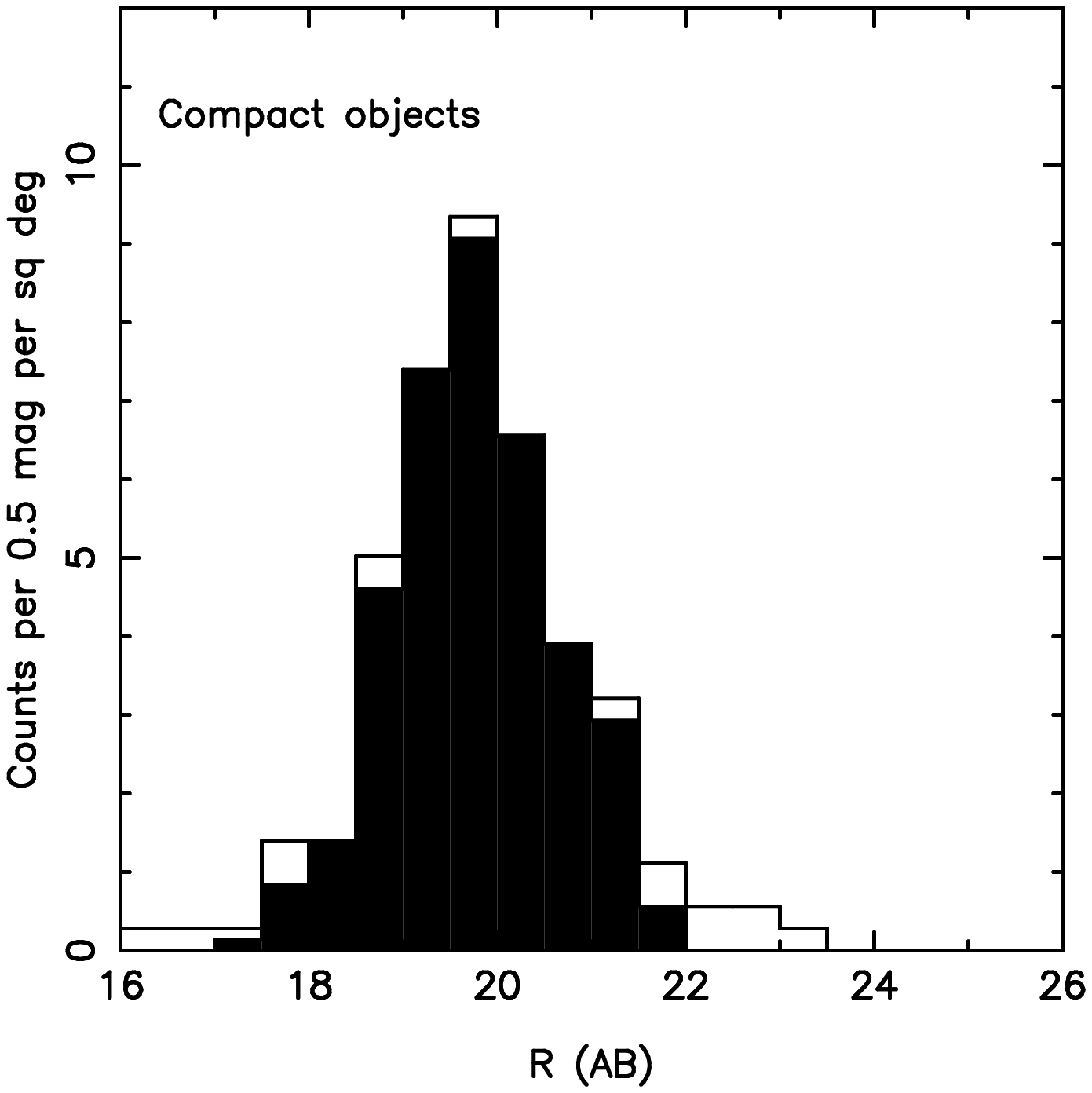}
\epsscale{1.00}
\caption{The distribution of $R$-band magnitudes for all
({\it left}) and optically compact ({\it right}) $24~\micron$
sources. The distribution of objects with spectroscopic redshifts (from both the
literature and AGES) is shown with the shaded histogram. Most (86\%) of objects with
compact optical morphologies have AGES spectroscopic redshifts.
\label{fig:dndm}}
\end{figure*}

Table~\ref{table:counts2} presents the number counts of the $24~\micron$ sources 
and Figure~\ref{fig:dndm} shows the distribution of their $R$-band magnitudes.
Most of the $24~\micron$ sources are brighter than $R=22$, and most 
of these sources are either quasars or $z \lesssim 1$ galaxies.  
The 17\% of $24~\micron$ sources with $R>21.7$ counterparts include include ultra-luminous     
infrared galaxies and obscured quasars at $z>1$ \markcite{hou05,mar05}({Houck} {et~al.} 2005; {Mart{\'{\i}}nez-Sansigre} {et~al.} 2005).
Type I quasars have relatively bright optical counterparts
and mostly compact optical morphologies. AGES spectroscopic redshifts are 
available for 63\% of the $S_{24}>1$~mJy sources and 92\% of the $24~\micron$ sources 
satisfying the AGES selection criteria.                                                        

\section{THE TYPE I QUASAR SAMPLE}
\label{sec:t1}

\begin{figure*}
\plotone{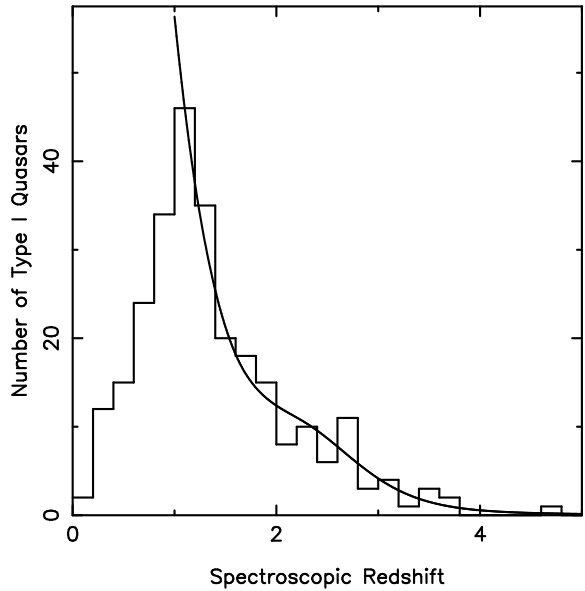}
\caption{The spectroscopic redshift histogram of
type I quasars. The sample is incomplete at $z<1$ as some
quasars are classified as extended in the deep NDWFS imaging.
The curve is derived from the best-fit luminosity function,
and is a good approximation to the $z>1$ data.
\label{fig:dndz}}
\end{figure*}

We define our type I quasar sample to be $24~\micron$ sources brighter
than $1~{\rm mJy}$ with $17.2<R<23.5$ compact optical counterparts.
The compact optical morphology is due to the observed optical emission
being dominated by a relatively unobscured AGN.
Our sample is incomplete for $z<1$ quasars and Seyferts, where
AGN host galaxies are sometimes resolved in the NDWFS imaging
and AGN host galaxies can contribute significantly to the observed $24~\micron$ flux. 
The bluest spectroscopic quasar in our sample has a color of $R-[24]=2.5$ 
(where $R-[24]$ is the apparent color in $AB$ magnitudes), 
so we have excluded  $R-[24]<2$ sources from the sample with the expectation that
this will not effect the sample completeness for highly luminous quasars. 
Of the 47 objects excluded with this color criterion, 90\% are brighter than                       
$R=17$ and are almost certainly stars. AGES spectroscopy identified 20 $z<1$ galaxies              
which were erroneously classified as compact in the NDWFS, and they were reclassifed
as extended and excluded from the quasar sample. 
Two $R<20.2$ $z>1$ quasars which were classified as                                        
extended in the optical imaging have been added to the quasar sample. 
The final sample contains 292 quasars and quasar candidates.
Figure~\ref{fig:dndz} shows the redshift distribution of the 270                                   
objects in the sample with AGES spectroscopic redshifts.
All the quasars without redshifts are at $z<3.8$ (or are blends
with $z<3.8$ objects), as they all have $B_W$ detections indicating that 
the Lyman limit is within or blueward of our $B_W$ filter. The vast majority (83\%)
of the quasars with spectra and 59\% of the candidates without spectra                              
have one or more photons in the X\bootes~Chandra imaging survey \markcite{mur05}({Murray} {et~al.} 2005).
In contrast only 23\% of the $z<1$ galaxies with $S_{24}>1~{\rm mJy}$ are detected in X\bootes.     
Quasar candidates without spectra have the same likelihood ($\simeq 10\%$) of having 
$1.4~{\rm GHz}$ counterparts \markcite{dev02}(from  {de Vries} {et~al.} 2002) as spectroscopically confirmed quasars. 
We therefore expect most of the 22 quasar candidates without spectra to be quasars.                 

\begin{figure*}
\plotone{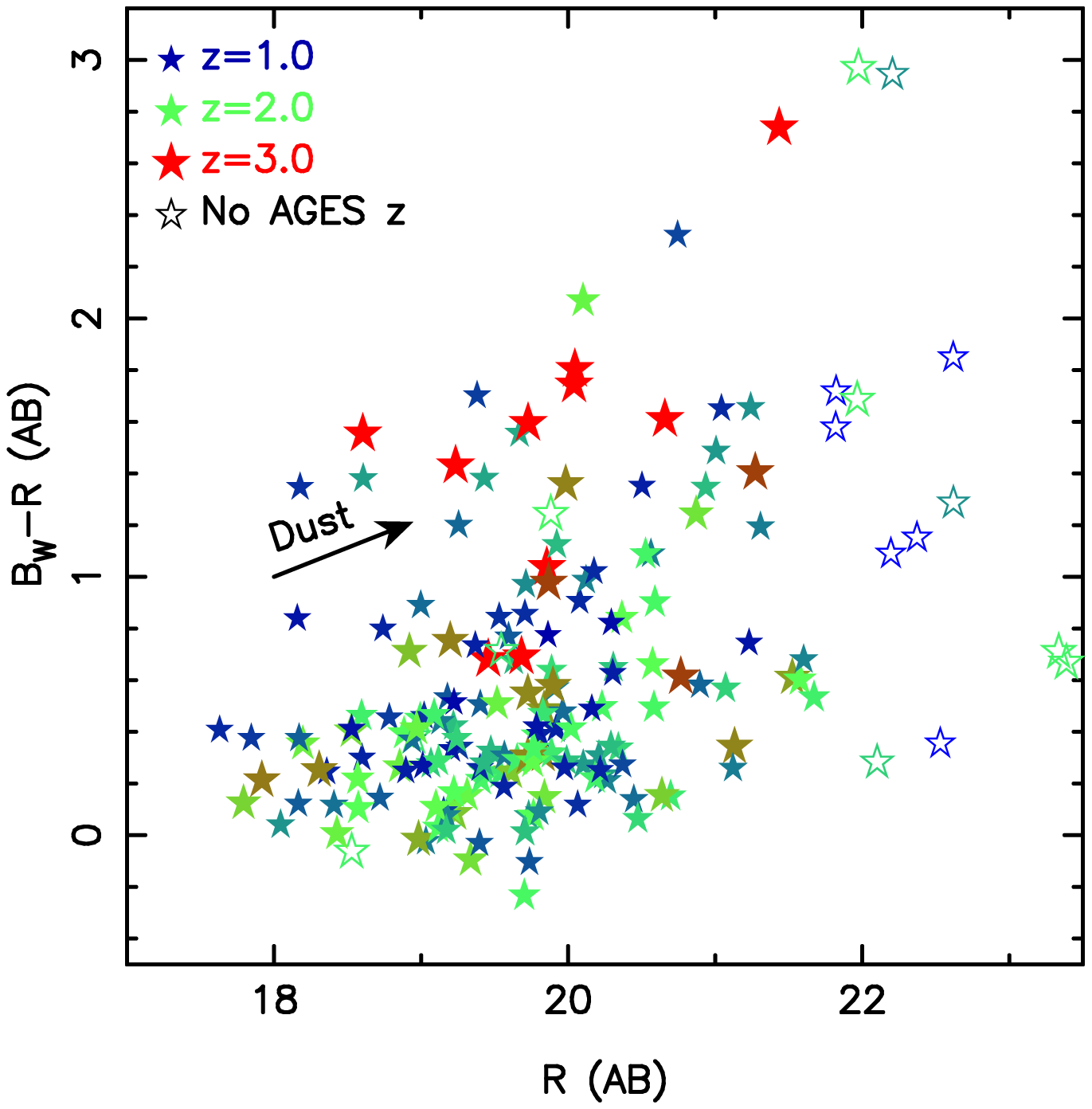}\plotone{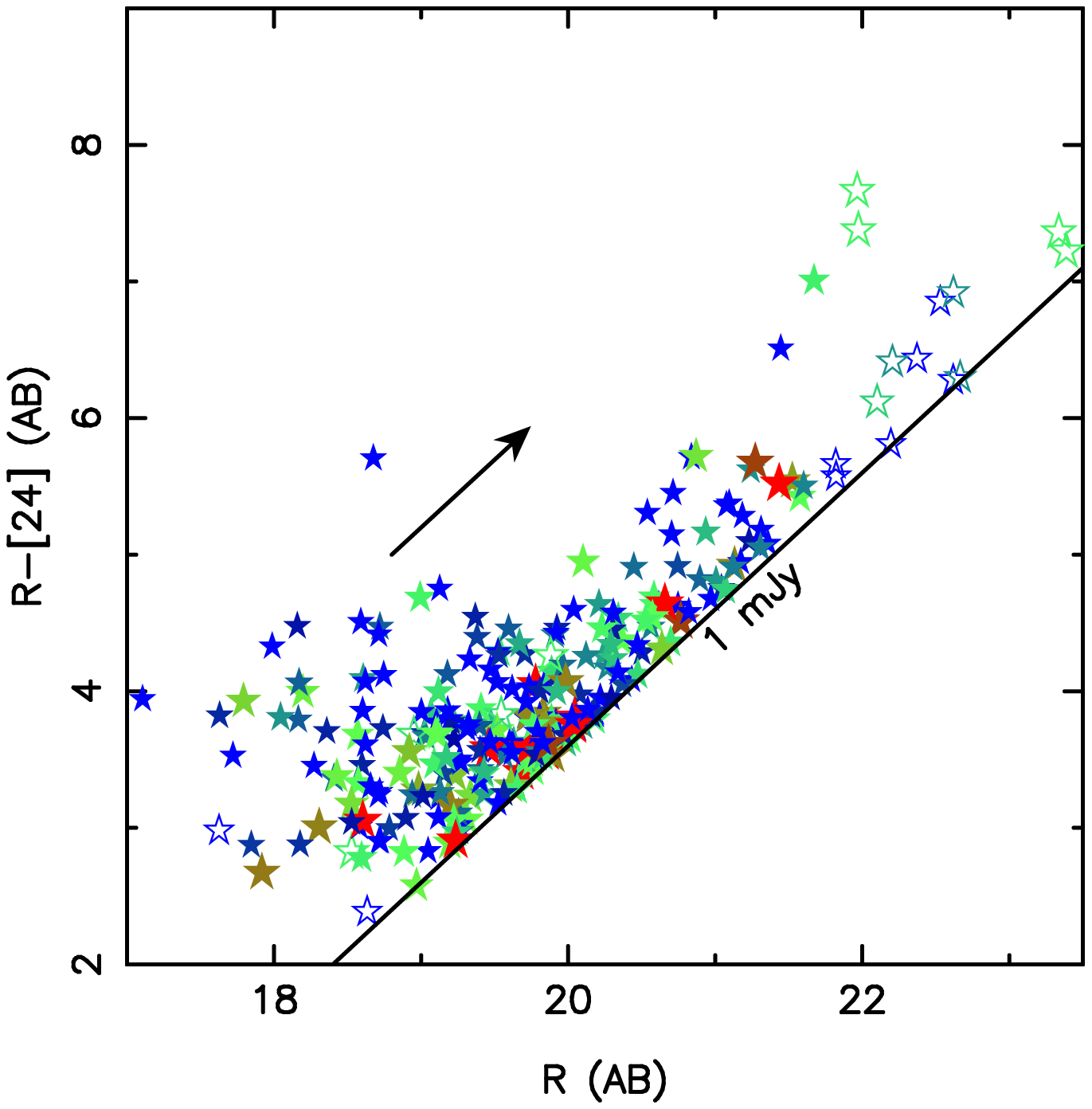}
\caption{The apparent colors of type I quasars.
The symbol size and color are a function of redshift.
Objects with spectroscopic and photometric redshifts are shown with
solid and open symbols respectively.
Arrows denote the apparent color vectors for $z=2$ dust extinction
with rest-frame $E(B-V)=0.1$, using the approximation of \markcite{cal99}{Calzetti} (1999).
Only objects with unsaturated $B_W$ photometry are shown in the left panel.
Relatively little dust obscuration can explain much of the dispersion of type I quasar
colors.
\label{fig:colr}}
\end{figure*}

While our $R$-band spectroscopic flux limit is less sensitive to 
extinction than standard ultraviolet excess (UVX) searches, it 
still corresponds to the rest-frame ultraviolet emission of the
quasars. As shown in Figure~\ref{fig:colr}, the apparent colors
and magnitudes of quasars show some evidence for obscuration
since fainter quasars tend to have redder colors that could
be explained by modest amounts of extinction.
This is consistent with modest dust reddening producing 
the broad red tail of type I quasar colors
\markcite{ric01}(e.g., {Richards} {et~al.} 2001).   

\begin{figure*}
\epsscale{0.95}
\plotone{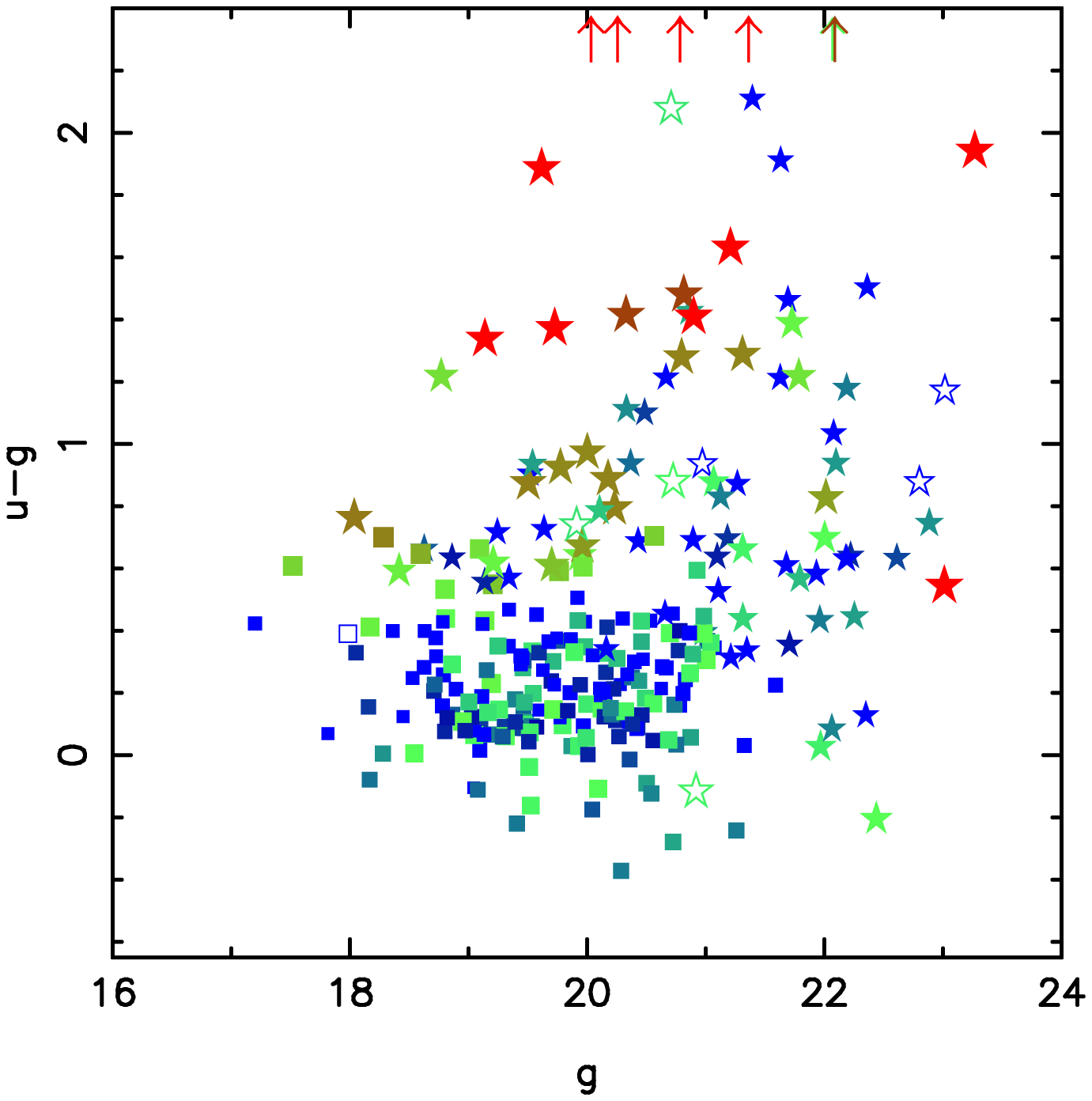}\plotone{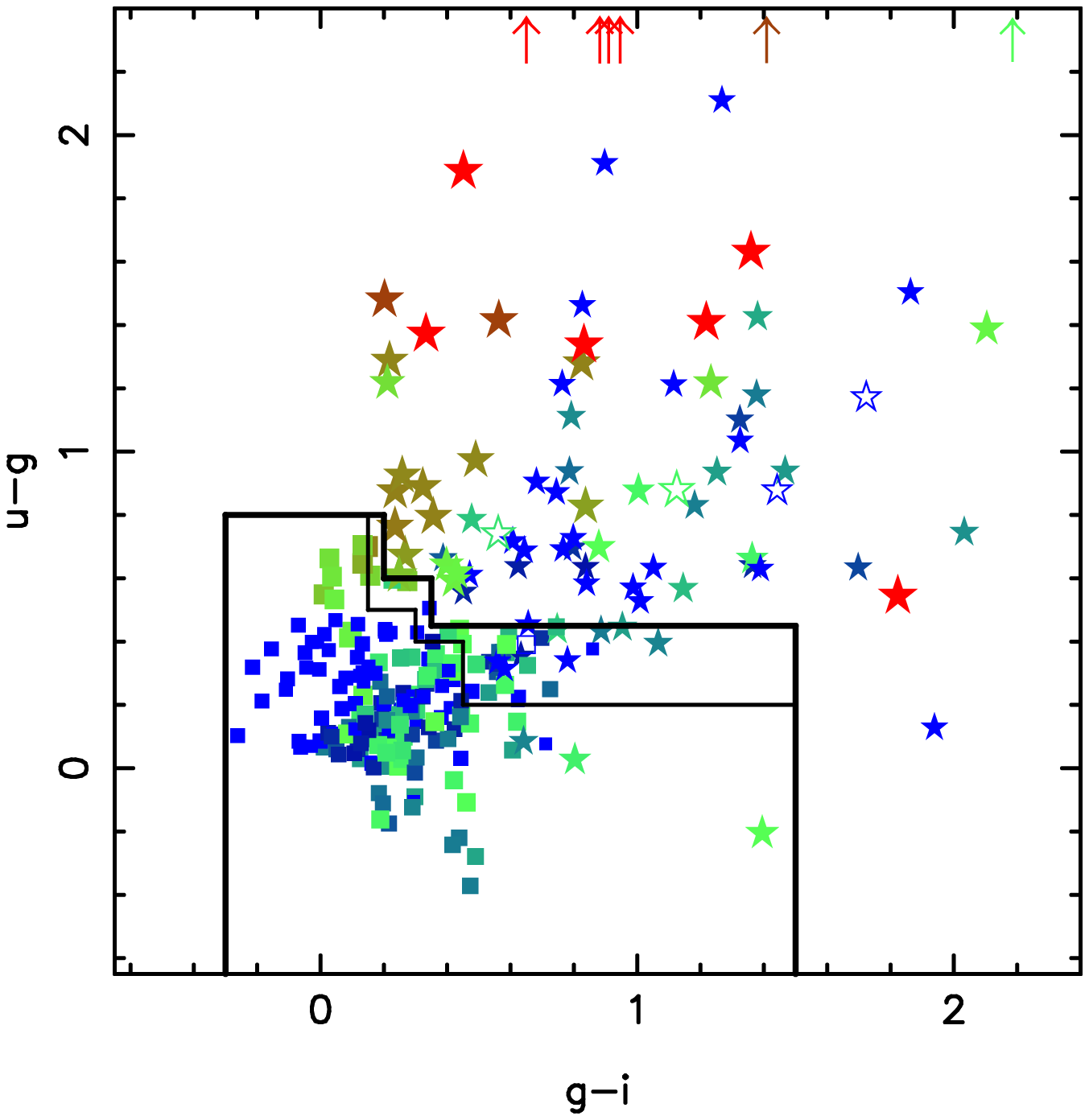}
\epsscale{1.00}
\caption{SDSS DR4 color-magnitude (left) and color-color (right) diagrams for the
$24~\micron$ selected type I quasar sample. As in Figure~\ref{fig:colr},
symbol size and color are a function of redshift, and open
symbols denote objects without spectroscopic redshifts. In the right panel,
the 2SLAQ selection criteria for  $g<21.15$ and $g<21.85$ are shown with
bold and narrow lines respectively. Quasars selected by the 2SLAQ criteria are
shown with squares. Of the 142 $1.0<z<2.2$ quasars in the $24~\micron$ sample,
112 (79\%) are also selected by the UVX criteria of 2SLAQ.                          
\label{fig:sdss}}
\end{figure*}

\subsection{Comparison with UVX Surveys}

As a check of our sample completeness and to compare our selection
criteria with traditional optical surveys, we compared our
sample to a sample of UVX quasar candidates. These were selected
using the criteria of the 2dF-SDSS Luminous Red Galaxy and Quasar 
Survey \markcite{ric05}(2SLAQ; {Richards} {et~al.} 2005) and SDSS DR4 imaging catalogs, which 
overlap almost all of the \bootes~ field. The 2SLAQ criteria do not include morphology
cuts for most objects, but do include optical color cuts which limit the
ability of 2SLAQ to detected reddened and $z>2.2$ quasars.
We restricted the comparison to $R<19$ quasars because at fainter 
magnitudes the blue quasars will tend to have $S_{24}<1$~mJy (see Figure~\ref{fig:colr}). 
Of the 46 $R<19$ UVX candidates in \bootes, 38 (83\%) are in our                                     
$24\micron$ quasar sample. The remaining $R<19$ UVX candidates are
starburst galaxies or  $z<0.3$ quasars whose host galaxies 
are detected in the NDWFS imaging (and are classified as extended sources).                         
Our $24\micron$ quasar sample is therefore highly complete for type I quasars 
selected with the UVX technique, particularly at $z>1$ where quasar hosts are 
unresolved in the NDWFS imaging.

We now consider the completeness of UVX surveys for quasars selected via their $24~\micron$ flux. 
Of the 23 $R<19$ quasars with $1.0<z<2.2$ in our $24~\micron$ sample, 19 ($83\%$) are                
selected by the 2SLAQ UVX criteria. The four $1.0<z<2.2$ quasars not
selected by the 2SLAQ criteria all have red colors ($u-g>0.5$ and $R-[24]>3.7$).                     
Given the problems of small number statistics, this is consistent with the estimate of
90\% completeness for the UVX selected quasars from the 2QZ catalog
by \markcite{smi05}{Smith} {et~al.} (2005).  However, as shown in Figure~\ref{fig:sdss}, the 
completeness of the UVX samples of quasars above a fixed infrared flux is lower.  
This is relevant for understanding the accretion history of quasars 
as the infrared is almost certainly more strongly correlated with the 
bolometric luminosity than the ultraviolet.
If we select the 142 $1.0<z<2.2$ quasars brighter than $1~{\rm mJy}$ at $24~\micron$,                
we find 79\% are selected by the 2SLAQ criteria. The quasars not selected by 
the 2SLAQ criteria are sometimes optically blue and faint, but generally they are
red. The median apparent colors of the 30 $1.0<z<2.2$ not selected by the 
2SLAQ criteria are $u-g=0.66$ and $R-[24]=4.62$. Clearly, there is a significant 
population of bolometrically luminous red type I quasars which are mostly undetected by traditional UVX 
quasar surveys. Red type I quasars comprise $\sim 20\%$ of the type I quasar population.

\subsection{Morphological Completeness}
\label{sec:mcomp}

We can also check the consequences of including only compact sources in our
sample.  This criterion certainly removes type II quasars from the sample,
as their optical emission is dominated by their host galaxies, but it will
also remove $z<1$ type I quasars from our sample when their host galaxies
are detected in the deep NDWFS images.  Thus, when we evaluate the quasar
luminosity function in \S\ref{sec:lf}, we restrict the analysis to $z>1$ quasars.  
AGES targeted both compact and extended $17.2<R<20.2$ sources, so for this magnitude
range we can estimate the incompleteness due to compact selection criteria.
Of 130 $24\micron$ sources with $z>1$ and $R<20.2$, only 
two quasars (1.5\%) were classified as extended sources.                            
These objects have been added to our spectroscopic quasar sample.
One is close to a foreground star while the other is a blend
with a galaxy, which may or may not be physically associated with the quasar.
At $20.2<R<21.7$, we assume the $z>1$ quasar sample is 1.5\% incomplete due to      
quasars being classified as extended in the optical imaging.
Thus, the morphology cut should only affect the sample completeness at 
levels well below our statistical uncertainties. 

\subsection{Spectroscopic Completeness}
\label{sec:scomp}

We can see in Figures~\ref{fig:dndm} and~\ref{fig:colr} that the vast
majority of $S_{24}>1~{\rm mJy}$ type I quasars are brighter than our 
spectroscopic limit of $R=21.7$ and have spectroscopic redshifts. 
The number of $S_{24}>1~{\rm mJy}$ type I quasars is clearly 
decreasing with increasing magnitude at $R>21.7$. Unlike UVX criteria, 
our selection criteria do not intrude into the main locus of type I quasar colors.
We can use a relatively simple model of the spectroscopic completeness, and
our estimate of the luminosity function is not particularly sensitive to 
changes in this model.

\begin{figure*}
\epsscale{0.85}
\plotone{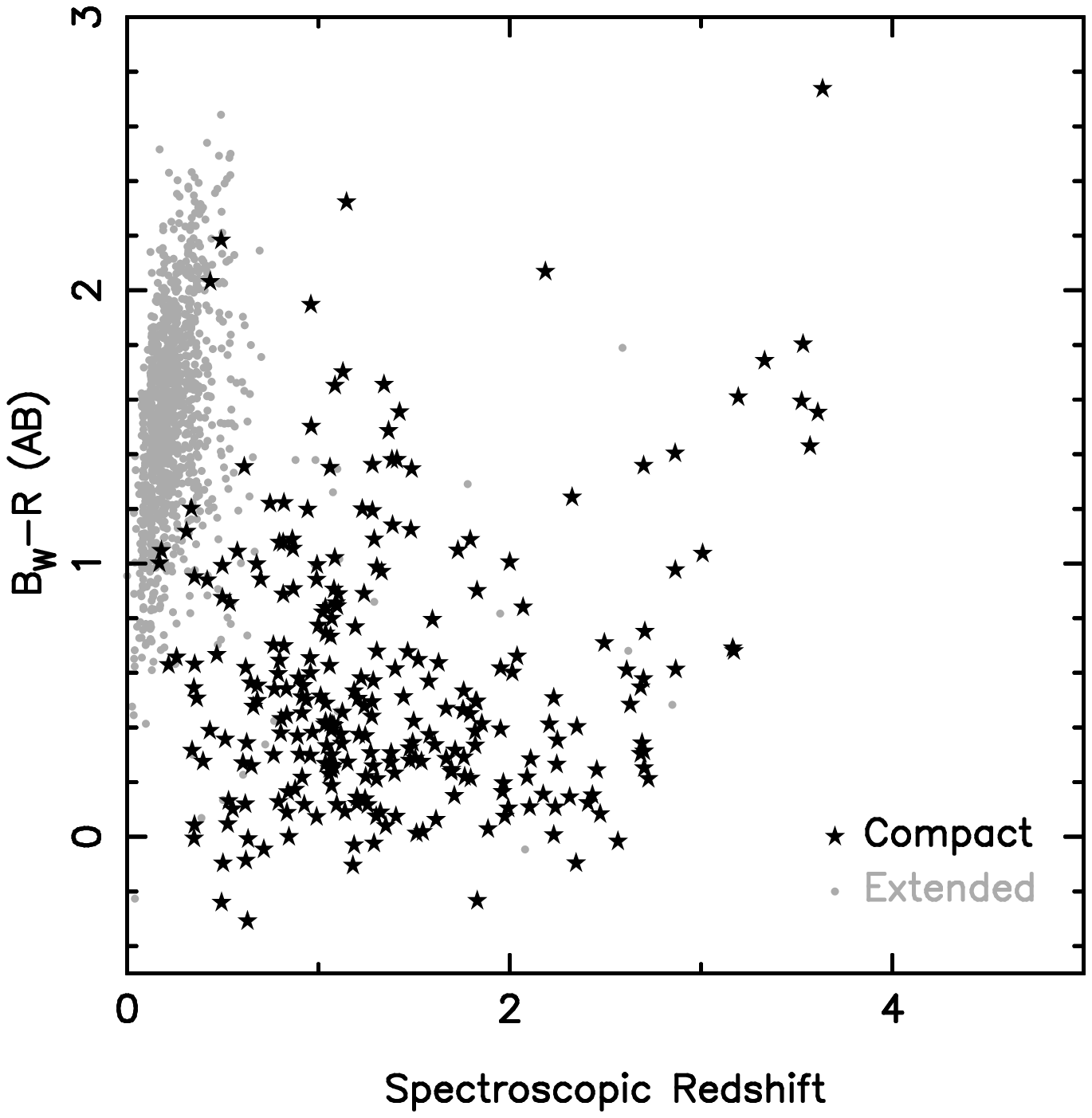}\plotone{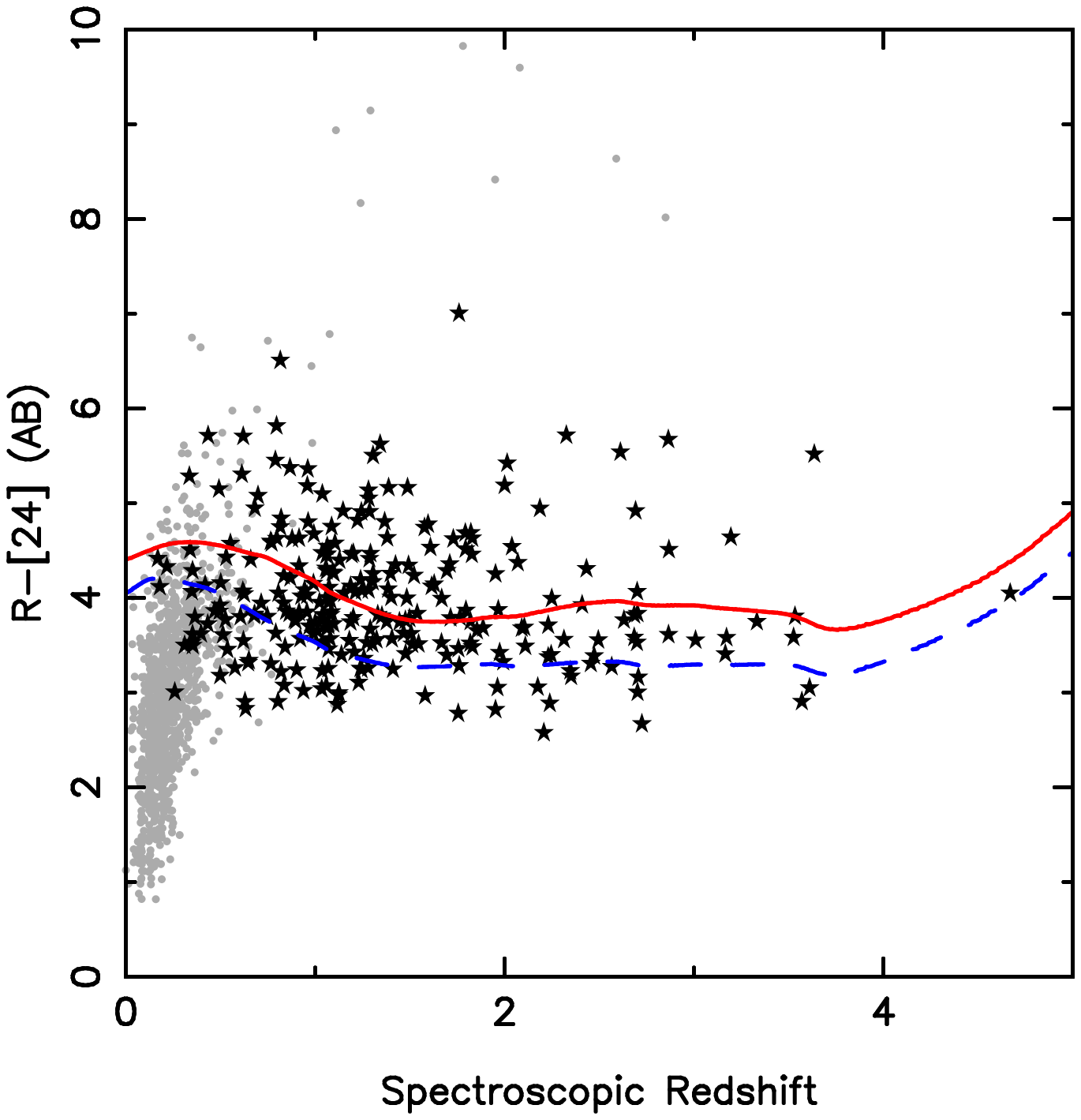}
\epsscale{1.00}
\caption{The apparent colors of $24~\micron$ sources in the Bo\"otes field as a
function of spectroscopic redshift. Optically compact objects are shown with stars
while extended objects are shown with dots. Two quasar templates from \markcite{hat05}{Hatziminaoglou} {et~al.} (2005)
are also shown. The dashed line denotes the template derived from 35 SDSS quasars
in the SWIRE ELAIS-N1 field while the solid line denotes the higher MIR flux quasar
template, which is a better fit to the median color of our sample. The apparent $R-[24]$
color of type I quasars is a weak function of redshift in the range $1<z<5$.
The spectroscopic sample contains very red and bolometrically luminous quasars,
though at the limits of the $24~\micron$ and AGES spectroscopic samples we are
restricted to $R-[24]<5.3$ objects.
\label{fig:colz}}
\end{figure*}

To model the spectroscopic completeness, we assume the 
distribution of rest-frame $R-[24]$ colors for $z>1$ type I quasars 
is neither a function of luminosity or redshift. We measure 
this distribution using the observed colors of $z>1$ quasars and $k$-corrections 
derived from the higher mid-infrared (MIR) flux quasar spectral energy 
distribution (SED) of \markcite{hat05}{Hatziminaoglou} {et~al.} (2005). When spectroscopic redshifts are 
unavailable, we estimate photometric redshifts by searching 
for the nearest neighbor in color-space with a spectroscopic redshift. 
As shown in Figure~\ref{fig:colz}, the apparent $R-[24]$ colors of quasars 
are a weak function of redshift so relatively inaccurate photometric redshifts 
can be used to determine rest-frame colors (but not luminosities).

\begin{figure*}
\plotone{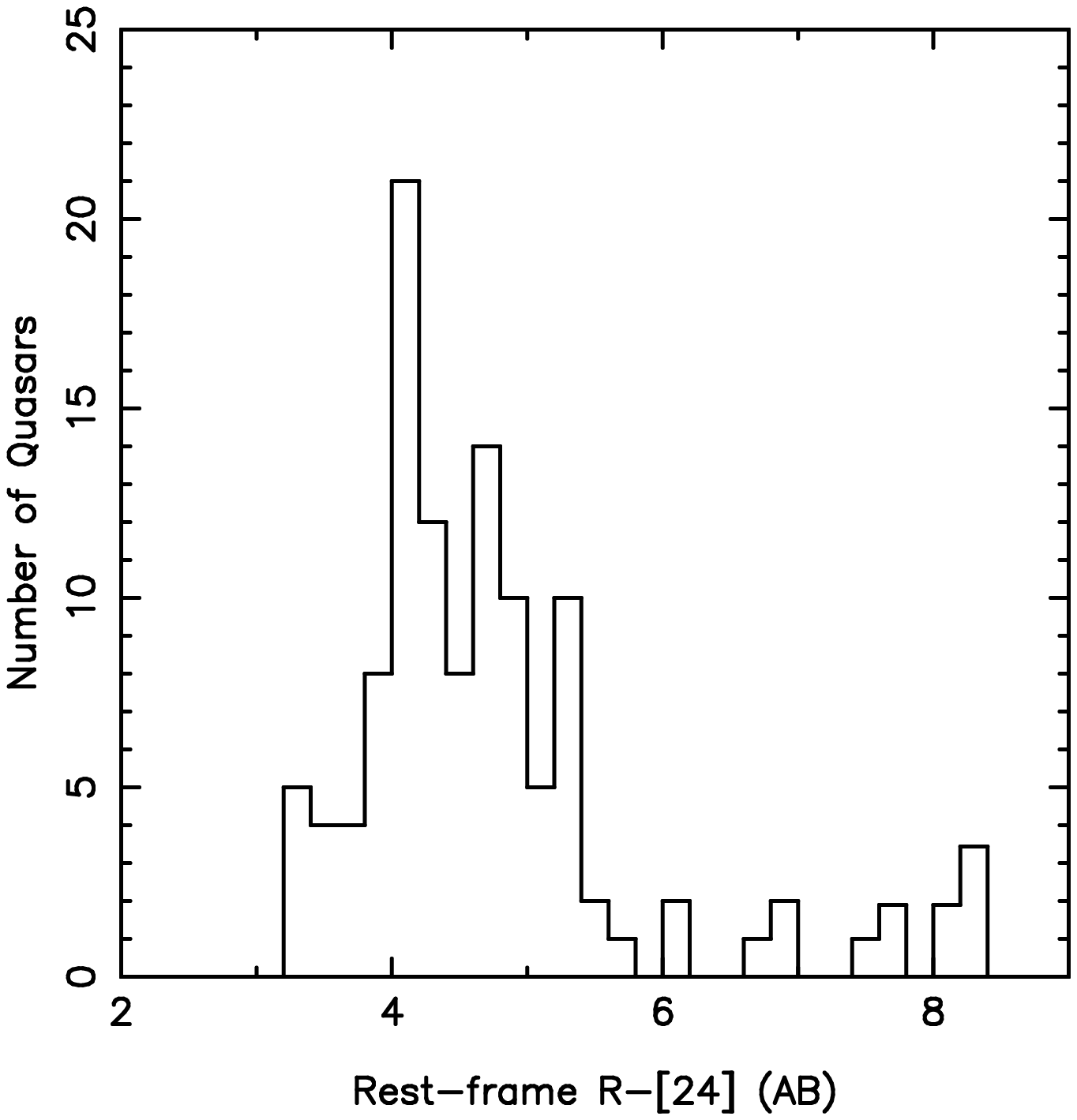}\plotone{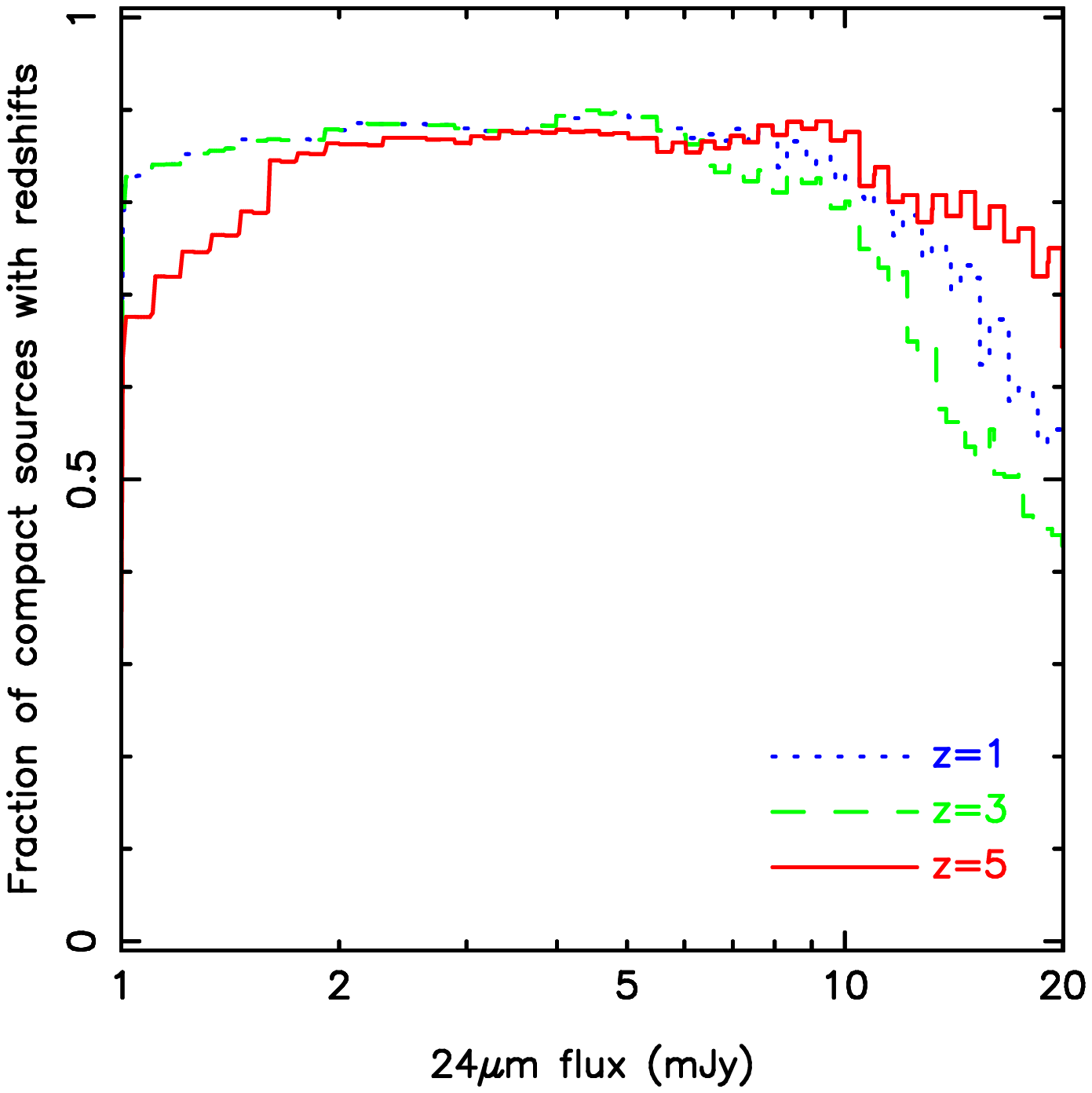}
\caption{The distribution of rest-frame colors (left panel) and the
estimated spectroscopic completeness (right panel) for the  $24~\micron$
selected type I quasar sample.
For the redshift range $1<z<4$, all $S_{24}>1~{\rm mJy}$ type I quasars with restframe $R-[24]<6$
are brighter than our $R=21.7$ spectroscopic limit. The plateau of the spectroscopic
completeness corresponds to $17.2<R<20.2$, where the AGES spectroscopic completeness
for optically compact sources is $\simeq 95\%$.
\label{fig:comp}}
\end{figure*}

Type I quasars fainter than $R=23.5$ are not included in our sample, 
as they are not classified as compact sources. We include a completeness 
correction for these objects by assuming their color distrbution is 
identical to quasars with higher $24~\micron$ fluxes and that the shape of 
their  $24~\micron$ number count  distribution is identical to that of bluer quasars.
We estimate there are four $z>1$ type I quasars fainter than $R=23.5$, and include
this in our completeness estimates though this has little effect on our measured
luminosity function.

Our estimate of the rest-frame $R-[24]$ distribution for $z>1$ quasars
is shown in the left panel of Figure~\ref{fig:comp}.
To derive the distribution of apparent $R$-band magnitudes as a function
of both $24~\micron$ flux and redshift, we used this model and applied
the $k$-corrections derived from the  higher MIR quasar SED of \markcite{hat05}{Hatziminaoglou} {et~al.} (2005).
We multiplied the distribution of $R$-band magnitudes by the measured 
completeness of as a function of $R$-band magnitude to determine the 
spectroscopic completeness of the sample as a function of 
$24~\micron$ flux and redshift.  As shown in the right panel of 
Figure~\ref{fig:comp}, the spectroscopic completeness peaks at a 
$24~\micron$ flux of $\sim 7 {\rm mJy}$, which corresponds 
to $17.2<R<20.2$ where the AGES spectroscopic completeness plateaus at 
$\simeq 95\%$.  
The spectroscopic completeness then declines to $\sim 80\%$ at 
both $1~{\rm mJy}$ (because of the unmeasured redshifts) and at 
$15$~mJy (because of the AGES $R>17.2$ criterion).

\section{LUMINOSITY FUNCTION OF TYPE I QUASARS}
\label{sec:lf}

The luminosity function of quasars is one of the principal constraints on the
accretion history of the most massive blackholes \markcite{kau00}(e.g., {Kauffmann} \& {Haehnelt} 2000).
At the highest luminosities the fraction of type II quasars is thought to be low
\markcite{ued03,hao05}(e.g., {Ueda} {et~al.} 2003; {Hao} {et~al.} 2005), although current studies do not include radio-quiet 
quasars obscured in both the optical and X-ray. If there are few type II quasars at
very high luminosities then the type I quasar luminosity function  
provides a good approximation of the overall quasar space density at
the highest luminosities.

We evaluate the luminosity function using 183 $z>1$ quasars with spectroscopic redshifts.
As discussed earlier, the deep NDWFS images can sometimes detect the host galaxy
of $z<1$ quasars so modeling the completeness would require a detailed 
treatment of star/galaxy classification as a function of quasar luminosity,
host luminosity, host morphology and image quality.  Several $z<1$ extended objects 
are classified as broadline AGNs by the AGES pipeline and are not included
in our sample of 292 type I quasars. The present sample does not include
$z>5$ quasars because of the bright $R$-band magnitude limit.
In the 2005 AGES sample, which uses a fainter $I$-band selection criterion, appreciable
numbers of $z>5$ quasars are found, but they are fainter than $S_{24}=1~{\rm mJy}$. 

\begin{figure}[hbt]
\plotone{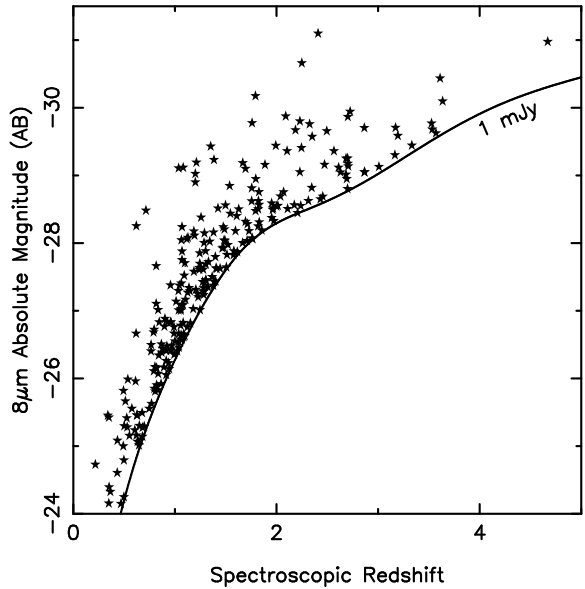}
\caption{The distribution of the $8~\micron$ absolute magnitudes of the type I
quasar sample as a function of spectroscopic redshift. Absolute magnitudes were
determined using the measured $24~\micron$ fluxes, spectroscopic redshifts, and
the higher MIR flux template of \markcite{hat05}{Hatziminaoglou} {et~al.} (2005).
Since rest-frame $8~\micron$ roughly corresponds to observed $24~\micron$,
we have measured $8~\micron$ absolute magnitudes and the
$8~\micron$ luminosity function.
The $M_B<-22.3$ criterion for quasars \markcite{ver03}({V{\' e}ron-Cetty} \& {V{\' e}ron} 2003)
roughly corresponds to an $8~\micron$ $AB$ absolute magnitude limit of $-25.3$.
\label{fig:absz}}
\end{figure}

We estimated the rest-frame $8~\mu{\rm m}$ absolute magnitudes of the sample
using the AGES spectroscopic redshifts, the $24~\mu{\rm m}$ flux,
and the higher MIR flux quasar SED of \markcite{hat05}{Hatziminaoglou} {et~al.} (2005).
For the redshift range $1<z<5$, observed $24~\micron$ corresponds to
rest-frame $12~\micron$ to $4~\micron$, so our $8~\mu{\rm m}$ absolute
magnitudes are not very sensitive to our choice of quasar SED.
Using a bluer quasar SED template, derived from 35 SDSS quasars
in the SWIRE ELAIS-N1 field \markcite{hat05}({Hatziminaoglou} {et~al.} 2005), alters the absolute
magnitudes of individual quasars by only a few tenths of a magnitude, 
our luminosity function parameters by $\lesssim1\sigma$, and our conclusions
little. The distribution of absolute magnitudes as a function of redshift is shown
in Figure~\ref{fig:absz}. Most objects have $8~\mu{\rm m}$ absolute magnitudes
brighter than $-25$, which corresponds to $1.6\times 10^{44} {\rm ergs}~{\rm s}^{-1}$
in $\nu F_\nu$.

\begin{figure*}[hbt]
\epsscale{1.25}
\plotone{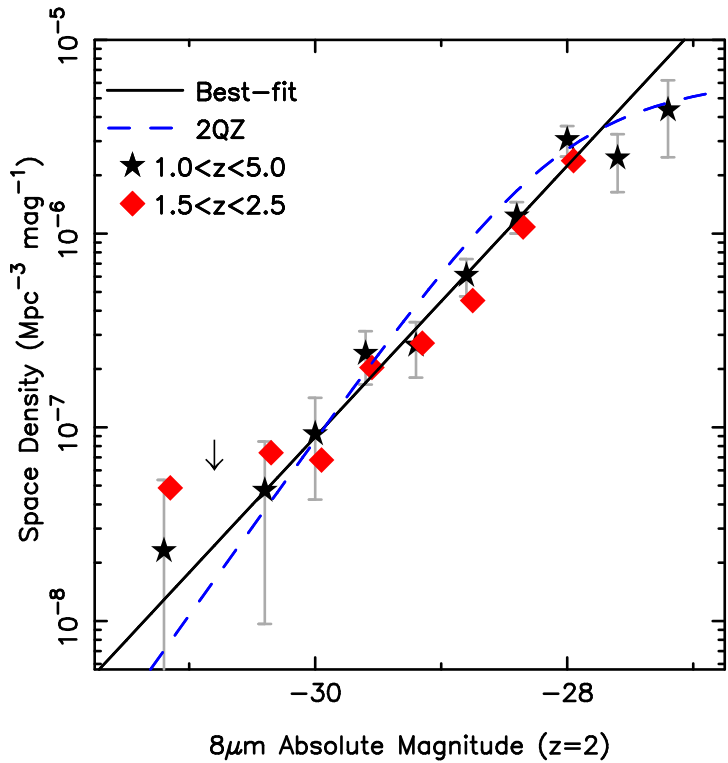}
\epsscale{1.0}
\caption{The $8~\micron$ luminosity function of $z=2$ type I quasars.
The solid line is the best-fit power-law while the datapoints
are determined using $1/V_{max}$ with quasar luminosities adjusted
for evolution. For clarity, we only show error bars and $3\sigma$ upper limits
for the $1/V_{max}$ luminosity function determined with $1<z<5$ quasars.
For comparison, the dashed line shows the 2QZ luminosity function
\markcite{cro04}({Croom} {et~al.} 2004) where we have assumed that $B_J-[8]=3.08~({\rm AB})$
for blue quasars. An $AB$ absolute magnitude of $-28.5$ corresponds
to $4.11\times 10^{45} {\rm ergs}~{\rm s}^{-1}$ in $\nu F_\nu$.
\label{fig:lf}}
\end{figure*}

We estimate the luminosity function using the maximum likelihood method of
\markcite{mar83}{Marshall} {et~al.} (1983).  We assume the luminosity function is a power-law of the form 
\begin{equation}
\phi_8 (L, z) dL \propto \left( \frac{L}{L^*(z)} \right)^{\alpha} dL
\end{equation}
where $\alpha$ is a constant and $L^*(z)$ describes the pure luminosity evolution of the
quasar luminosity function. This can also be written as a function of absolute magnitude;
\begin{equation}
\phi_8 (M, z) dM \propto 10^{0.4(\alpha+1)[M^*(z)-M]} dM.
\end{equation}
Optical surveys generally require models where the slope of the luminosity
function becomes shallower for low luminosity quasars \markcite{sch83,boy88,cro04,wol03,ric05}(e.g., {Schmidt} \& {Green} 1983; {Boyle}, {Shanks}, \& {Peterson} 1988; {Croom} {et~al.} 2004; {Wolf} {et~al.} 2003b; {Richards} {et~al.} 2005),
but a single power-law fits our data reasonably well. When we tried fitting
a double power-law to the data, we were unable to precisely constrain the faint-end 
slope due to the limited size and luminosity range of our sample. 
We modeled the evolution of the quasar luminosity function using
pure luminosity evolution where
\begin{equation}
L^*(z)=L^*(0) 10^{k_1 z + k_2 z^2 + k_3 z^3}
\end{equation}
with $k_1$, $k_2$, and $k_3$ being constants.  This is similar to the 
polynomial used by optical quasar surveys such as the 2QZ, and aids 
comparisons with the previous surveys. Unlike optical quasar samples,
we have no difficulty finding quasars at $2.3<z<3.0$ (where the
optical colors match those of main sequence stars), so we can track
the evolution of the luminosity function through its peak.  This
forces us to add the $k_3z^3$ term to the polynomial so that
the evolution on the high redshift side of the peak is not forced
to be identical to that on the low redshift side of the peak
by the symmetry of the equation.  This pure luminosity evolution
model is sufficient for a sample of $\sim 10^2$ quasars -- larger 
samples containing $\sim 10^4$ quasars require more complex 
models \markcite{cro04}(e.g., {Croom} {et~al.} 2004). 

\begin{figure*}[hbt]
\epsscale{1.10}
\plotone{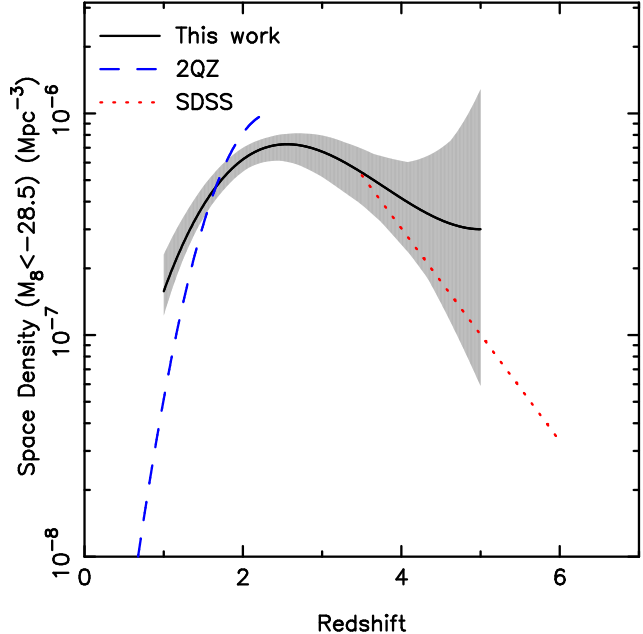}
\epsscale{1.0}
\caption{The space density of type I quasars brighter than $M=-28.5$ ($4.11\times 10^{45} ~{\rm ergs}~{\rm s}^{-1}$)
at $8~\micron$. The shaded region denotes the area bounded by the $\pm 1\sigma$ uncertainties. We have overplotted space
densities from the 2QZ \markcite{cro04}({Croom} {et~al.} 2004) and SDSS \markcite{fan04}({Fan} {et~al.} 2004) by assuming that the \markcite{hat05}{Hatziminaoglou} {et~al.} (2005) template derived
from 35 SDSS quasars approximates the median color for these samples.  This is only an approximation since
it ignores the spectral diversity of blue quasars.  The space density of bright quasars rapidly
evolves and peaks at $z=2.5\pm 0.3$.
\label{fig:lfz}}
\end{figure*}

In Figure~\ref{fig:lf} we plot our best-fit luminosity function
for $z=2$ type I quasars and our best-fit parameters are
summarized in Table~\ref{table:quasarlf}. Only the 183 quasars                
with spectroscopic redshifts of $z>1$ were used to measure the 
luminosity function. We used the completeness estimates described in \S\ref{sec:mcomp} and \S\ref{sec:scomp} to 
correct for the spectroscopic and morphological incompleteness of the sample.
We also generated an estimate of the $z=2$ luminosity function
using the $1/V_{max}$ method after attempting to remove the effects
of evolution.  For each quasar, we estimated a $z=2$ absolute magnitude by assuming 
\begin{eqnarray}
M(2) & = & M(z) +  2.5 ( k_1 z + k_2z^2 + k_3 z^3) \nonumber \\
     &   & - 2.5 ( 2 k_1 + 4 k_2 + 8 k_3) 
\end{eqnarray}
with our best-values of $k_1$, $k_2$, and $k_3$ (Table~\ref{table:quasarlf}).
This allows us to use quasars from a broad range of redshifts to estimate 
a $1/V_{max}$ luminosity function.  We have used a statistical representation of 
the evolution rather than a model for the evolution of individual quasars.
A large number of black holes undergoing relatively brief episodes of quasar 
activity can produce a luminosity function that mimics a smaller population of quasars undergoing 
luminosity evolution \markcite{kau00}(e.g., {Kauffmann} \& {Haehnelt} 2000). Our measurements of the 
$z=2$ quasar luminosity function using the  $1/V_{max}$ method are plotted 
in Figure~\ref{fig:lf} and tabulated in Table~\ref{table:binned}.
Our best fit power-law and $1/V_{max}$ luminosity functions agree well.
The very faintest quasars show a slight departure from the power-law, 
but we cannot detect it with any precision when we 
include a break in our maximum likelihood models. Figure~\ref{fig:dndz} 
shows that the observed and model distribution of the $1<z<5$ quasars in 
redshift agree well, though our results cannot be naively extrapolated to other
redshifts.

Our best-fit parameters for the quasar luminosity function are 
summarized in Table~\ref{table:quasarlf} along with the 
luminosity function parameters for the 2QZ \markcite{cro04}({Croom} {et~al.} 2004) for
comparison.  Our luminosity function is not as steep as the bright-end
of the 2QZ luminosity function. However, we are measuring quasars
that are slightly brighter than the break in the 2QZ luminosity function,
where the power-law index changes from $-3.3$ to $\simeq -1.5$.
As shown in Figure~\ref{fig:lf}, the $z=2$ luminosity functions of this work
and the 2QZ are similar for quasars within our luminosity range.
Figure~\ref{fig:lfz} plots the measured evolution of the quasar space
density from our $24~\micron$ quasar sample and the optically
selected 2QZ and SDSS samples. The three surveys indicate a peak in 
quasar activity occurs at $z\simeq 2.5$.

If the dust obscuration of AGNs is a function of bolometric luminosity 
\markcite{hao05}(e.g., {Hao} {et~al.} 2005), then the shape of the quasar luminosity function should
vary with color.  For example, if obscuration increases with decreasing 
luminosity, then the luminosity function of reddened quasars should be
steeper than that of blue quasars.  We divided our sample into quasars 
with $R-[24]$ colors that are redder and bluer than the
higher MIR flux quasar template of \markcite{hat05}{Hatziminaoglou} {et~al.} (2005). This rest-frame color cut of $R-[24]=4.4$ corresponds to an 
apparent color cut of $R-[24]\sim3.8$ over the redshift range $1<z<4$. 
As the spectroscopic subsamples contain $\simeq 90$ quasars 
each, the uncertainties for $\alpha$ are prohibitively large if we let evolution 
parameters ($k_1$, $k_2$, and $k_3$) float. We have therefore fixed the values of 
the evolution parameters to the fit derived with the entire quasar sample. 

Our best fit luminosity function parameters are given in Table~\ref{table:quasarlf}.
The power-law indices ($\alpha$) of the blue and red subsamples differ by only $1\sigma$.
This is not completely unexpected as the range of dust obscuration probed by our samples
is limited. As shown in Figure~\ref{fig:colr}, at $z=2$ our $17.2<R<21.7$
spectroscopic sample spans a range of $E(B-V)$ values of $\sim 0.4$.
In contrast, the range of $E(B-V)$ values spanned by Seyfert Is and IIs
is $\sim 100$ \markcite{cla00}(e.g., {Clavel} {et~al.} 2000). At present, the dependence
of the dust obscuration on bolometric luminosity has only been seen in 
comparisons of type I and II AGNs \markcite{hao05}({Hao} {et~al.} 2005), so a  
complete sample of type II quasars may be required to understand the 
relationship between dust obscuration and luminosity for the most bolometrically 
luminous quasars.

\section{TYPE II QUASARS}
\label{sec:obscured}

Ideally we would measure the luminosity functions of both type I 
and II quasars. However, spectroscopic redshifts for 
$z>1$ type II quasars are non-trivial as their optical counterparts
can be extremely faint \markcite{hou05,mar05}(e.g., {Houck} {et~al.} 2005; {Mart{\'{\i}}nez-Sansigre} {et~al.} 2005). Photometric redshifts are also difficult 
as the broad-band SED includes unknown contributions from the AGN, host galaxy, and
dust \markcite{zak05}({Zakamska} {et~al.} 2005). In this section, we provide a brief
discussion of the type II quasar candidates. A spectroscopic campaign to measure the 
redshifts of the most luminous type II quasar candidates is ongoing, and a future paper 
will discuss the luminosity function of these objects in detail.

We designate all $24~\micron$ sources brighter than $1~{\rm mJy}$ with 
extended $R>21.7$ optical counterparts (or counterparts too faint for reliable
morphological classification) as $z>1$ type II quasar candidates.
At $z\sim 1$, $R=21.7$ corresponds to a blue $\simeq L^*$ galaxy \markcite{wol03g}({Wolf} {et~al.} 2003a).
Unlike radio and X-ray surveys, we can detect $z>1$ type II quasars which 
are both radio-quiet and Compton thick. The main contaminants of the sample
will be starburst galaxies, $z<1$ type II AGNs, and reddened type I quasars. 
There are 376 $z>1$ type II quasar candidates in our survey area. Only 20 of   
these objects have spectroscopic redshifts, mostly from \markcite{hou05}{Houck} {et~al.} (2005), and 
all but one of these redshifts are in the range $0.7<z<3.0$.                                        

\begin{figure}[hbt]
\plotone{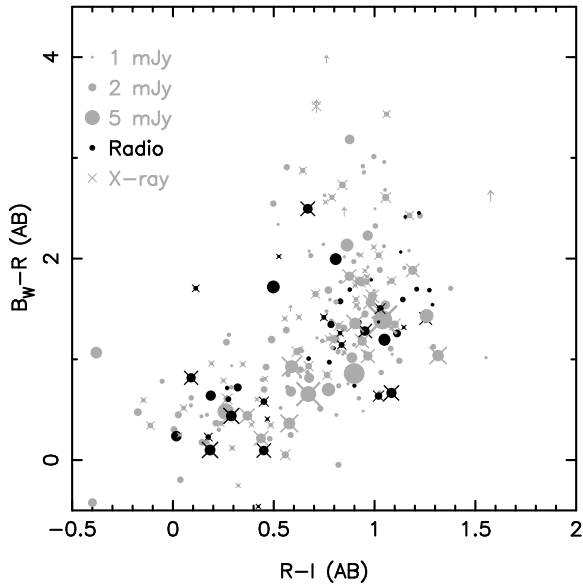}
\caption{The $B_WRI$ colors of $R>21.7$ type II quasar candidates.
Symbol size is a function of $24~\micron$ flux and objects with radio (bold symbols)
and X-ray counterparts (crosses) are highlighted.
The optical colors of most objects are consistent with
$1<z<3$ host galaxies. Candidates with X-ray counterparts are almost certainly AGNs, and they
comprise 32\% and 45\% of the candidates brighter than $1~{\rm mJy}$ and $2~{\rm mJy}$ respectively.
\label{fig:col_type2}}
\end{figure}

The optical colors of the type II quasars candidates are shown in Figure~\ref{fig:col_type2}.
The locus of blue $B_W-R$ and red $R-I$ colors are consistent with $1<z<3$ host galaxies.
Type II candidates with counterparts in the radio catalog of \markcite{dev02}{de Vries} {et~al.} (2002) and
the X\bootes~Chandra imaging survey \markcite{mur05}({Murray} {et~al.} 2005) are also shown in Figure~\ref{fig:col_type2}.
As the $0.5$ to $7~{\rm keV}$ background of the X\bootes~survey is only $\simeq 0.03$ counts within 
a $2^{\prime\prime}$ radius aperture, the detection of a single X-ray photon within this
aperture is significant in the vast majority of cases, and at $z=1$ a single photon 
in X\bootes~ corresponds to an X-ray luminosity of $\sim 10^{43}~{\rm ergs~s}^{-1}$.
For a $z>1$ source this is a strong indicator of AGN activity, as the implied luminosity 
is an order of magnitude more than low redshift starbursts including
Arp 220, which has an observed X-ray luminosity of $1.4\times 10^{41}~{\rm ergs~s}^{-1}$ \markcite{mcd03}({McDowell} {et~al.} 2003).
One-third of the type II quasar candidates have one or 
more X-ray photons in the Chandra images, and this increases to 45\% 
for candidates brighter than $2~{\rm mJy}$.                                                 

\begin{figure}[hbt]
\plotone{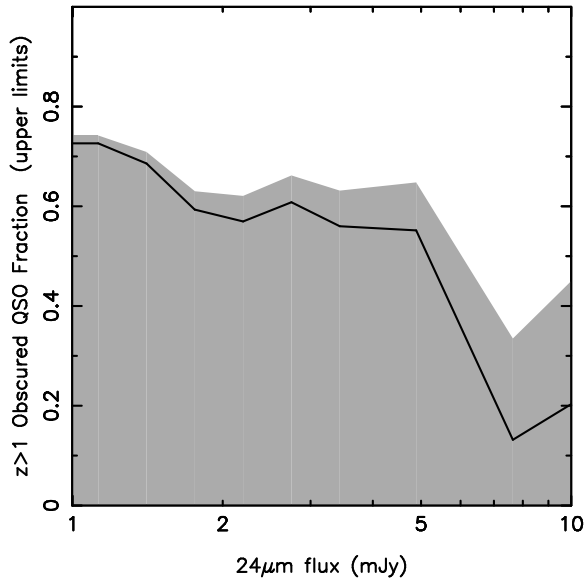}
\caption{Upper limits on the fraction of obscured quasars as a function of observed $24 \mu {\rm m}$ flux.
The shaded area denotes the region below our best estimate of the obscured fraction plus the $1\sigma$ statistical
uncertainties. Our sample of type II quasar candidates could be contaminated by starburst galaxies,
so we can only set upper limits in the absence of spectroscopic confirmation.
The fraction of obscured quasars could be as
high as 70\% at $1~{\rm mJy}$, and then decreases with increasing $24~\micron$ flux.
\label{fig:frac}}
\end{figure}

We have determined upper limits for the fraction of $z>1$ obscured 
quasars as a function of flux by dividing the number of $z>1$ type 
II candidates by the total number of $z>1$ type I and type II candidates. 
Contamination of our $z>1$ type I sample is low and contamination of the
$z>1$ type II candidate sample by $z<1$ AGNs and starbursts could be high, so 
this fraction is an upper limit. Upper limits for the fraction of obscured 
quasars as a function of $24~\mu {\rm m}$ flux are plotted Figure~\ref{fig:frac}.
This should not be confused with the fraction of obscured quasars
as a function of bolometric luminosity, which would require accurate 
redshift information. We do not know if the obscured quasars have 
systematically lower redshifts (and bolometric luminosities) than the 
type I quasars with comparable $24~\mu {\rm m}$ fluxes. We also do 
not know the extent of dust absorption of rest-frame $\sim 8~\mu{\rm m}$ 
emission in these AGNs. Despite these caveats, it is reasonable
to conclude that the ratio of obscured and unobscured AGNs near the break in 
optical luminosity function (roughly $S_{24}= 1~{\rm mJy}$) is of order unity.
This is consistent with studies of  X-ray selected AGNs, 
optically selected Seyferts, and small samples of infrared selected $z>1$ type II quasars 
\markcite{ste03,ued03,hao05,mar05}(e.g., {Steffen} {et~al.} 2003; {Ueda} {et~al.} 2003; {Hao} {et~al.} 2005; {Mart{\'{\i}}nez-Sansigre} {et~al.} 2005).
The fraction of obscured quasars decreases with 
increasing flux, which is similar to the trend observed as a function 
of luminosity in X-ray and optical samples \markcite{ued03,hao05}(e.g., {Ueda} {et~al.} 2003; {Hao} {et~al.} 2005). 
If the obscured fraction is decreasing rapidly with increasing luminosity, 
the type I quasar luminosity function approximates the luminosity
function of all quasars at the highest luminosities. However, to confirm this
will require complete spectroscopic samples of type II quasars.

\section{SUMMARY}
\label{sec:sum}

We have measured the $8~\micron$ luminosity function of type I quasars as a function 
of redshift using 292 quasars and quasar candidates selected from {\it Spitzer Space Telescope}        
$24~\micron$ imaging of the NDWFS \bootes~ field.  Spectroscopic redshifts for 270 of these quasars were 
measured as part of the AGES survey of \bootes. We were able to detect reddened type I 
quasars which were missed by UVX quasar surveys, and these comprise approximately a 
fifth of the type I quasar population.  
The type I quasar luminosity function at $8~\micron$, and its evolution
with redshift, is very similar to quasar luminosity functions at optical and ultraviolet wavelengths. 
Figure~\ref{fig:lfz} illustrates the evolution of the space density of type I quasars. 
There is qualitative agreement between our best-fit luminosity function 
and the 2QZ and SDSS, despite our sample being selected at $24~\micron$ rather 
than the optical. We directly detect the peak in quasar activity at 
$z=2.6\pm 0.3$, which has been difficult to directly measure                                           
with optical quasar surveys because it corresponds to the redshift 
range where quasar optical colors are similar to those of stars.

Our best-fit model indicates that the quasar space density 
declines by a factor of $\simeq 2$ between $z=2.5$ and $z=4.0$,
though this decline has low significance. If we fit our data
with a model where the space density is constant after 
$z_{peak}$ then our luminosity function parameters change by $\lesssim 1\sigma$. 
However, such a model disagrees with the results of optical surveys of $z>3$ quasars
\markcite{sch95,fan04}({Schmidt}, {Schneider}, \& {Gunn} 1995; {Fan} {et~al.} 2004).  Thus, unless the obscured quasar fraction increases
at high redshift, our study and previous optical quasar surveys indicate 
that the quasar space density declines at $z>3$.  

While the evolution of the type I quasar space density has been measured over a broad
range of wavelengths, the evolution of type II quasars remains uncertain.
Figure~\ref{fig:frac} indicates that they could comprise 70\% of $z>1$                                 
quasars brighter than $1~{\rm mJy}$, though this fraction decreases 
with increasing flux and luminosity. Until there is a systematic campaign
to measure the spectroscopic redshifts of type II quasars, the $8~\micron$
luminosity function of all quasars will remain uncertain.

\acknowledgments

We thank our colleagues on the NDWFS, MIPS, IRS, and AGES teams.
This paper would not have been possible without the efforts of the  KPNO and MMT  observing support staff.
We are grateful to Frank Valdes, Lindsey Davis and the IRAF team for the packages used to reduce the 
imaging data. We thank Alyson Ford, Lissa Miller, and Jennifer Claver, for
reducing much of the KPNO MOSAIC data used for this paper. 
Evanthia Hatziminaoglou kindly provided the quasar SED templates prior to publication.
Gordon Richards provided many useful suggestions with regards to the survey technique and manuscript.
This work is based in part on observations made with the {\it Spitzer Space Telescope}, which is operated 
by the Jet Propulsion Laboratory, California Institute of Technology 
under NASA contract 1407. This research was supported by the National Optical Astronomy Observatory which is
operated by the Association of Universities for Research in Astronomy (AURA), Inc.
under a cooperative agreement with the National Science Foundation.
Spectroscopic observations reported here were obtained at the MMT Observatory, 
a joint facility of the Smithsonian Institution and the University of Arizona.

\bibliography{}

\begin{deluxetable}{cccccccc}
\tablecolumns{8}
\tabletypesize{\scriptsize}
\tablecaption{Compact and extended object number counts for the multiwavelength sample.\label{table:counts2}}
\tablehead{
\colhead{$24~\micron$} &
\colhead{}    &
\multicolumn{6}{c}{Number counts} \\
\colhead{flux (mJy)}         &
\colhead{Total}       &
\colhead{AGES Spectra} &
\colhead{$R\leq17.2$}        &
\colhead{$17.2<R\leq 20.2$}  &
\colhead{$20.2<R\leq 21.7$}  &
\colhead{$21.7<R\leq 23$}  &
\colhead{$R>23.0$\tablenotemark{a}}                     
}        
\startdata
\cutinhead{Optically compact sources\tablenotemark{b}}
   1 - 50  &   337 &  268 &        45 &  197 &   81 &   12 &    2 \\
   1 -  2  &   222 &  181 &        23 &  120 &   67 &   10 &    2 \\
   2 -  3  &    54 &   46 &         5 &   40 &    9 &    0 &    0 \\
   3 -  4  &    21 &   16 &         5 &   13 &    3 &    0 &    0 \\
   4 -  5  &     7 &    6 &         1 &    5 &    1 &    0 &    0 \\
   5 -  6  &     8 &    3 &         2 &    4 &    1 &    1 &    0 \\
   6 -  7  &     6 &    4 &         1 &    4 &    0 &    1 &    0 \\
   7 -  8  &     4 &    4 &         0 &    4 &    0 &    0 &    0 \\
   8 -  9  &     3 &    2 &         1 &    2 &    0 &    0 &    0 \\
   9 - 10  &     0 &    0 &         0 &    0 &    0 &    0 &    0 \\
  10 - 20  &     9 &    5 &         5 &    4 &    0 &    0 &    0 \\
  20 - 30  &     2 &    1 &         1 &    1 &    0 &    0 &    0 \\
  30 - 40  &     1 &    0 &         1 &    0 &    0 &    0 &    0 \\
  40 - 50  &     0 &    0 &         0 &    0 &    0 &    0 &    0 \\
\cutinhead{Optically extended sources}
   1 - 50  &  1965 & 1181 &       232 & 1061 &  296 &  170 &  206 \\
   1 -  2  &  1322 &  735 &        86 &  700 &  229 &  138 &  169 \\
   2 -  3  &   324 &  224 &        41 &  201 &   37 &   17 &   28 \\
   3 -  4  &   128 &   87 &        27 &   74 &   12 &   11 &    4 \\
   4 -  5  &    65 &   43 &        17 &   33 &   10 &    0 &    5 \\
   5 -  6  &    33 &   27 &        10 &   18 &    3 &    2 &    0 \\
   6 -  7  &    20 &   13 &         7 &    9 &    4 &    0 &    0 \\
   7 -  8  &    15 &   14 &         6 &    9 &    0 &    0 &    0 \\
   8 -  9  &     9 &    6 &         4 &    3 &    1 &    1 &    0 \\
   9 - 10  &     8 &    6 &         3 &    5 &    0 &    0 &    0 \\
  10 - 20  &    24 &   15 &        15 &    8 &    0 &    1 &    0 \\
  20 - 30  &    12 &    7 &        12 &    0 &    0 &    0 &    0 \\
  30 - 40  &     3 &    2 &         3 &    0 &    0 &    0 &    0 \\
  40 - 50  &     2 &    2 &         1 &    1 &    0 &    0 &    0 \\
\enddata
\tablenotetext{a}{At faint magnitudes, where the SExtractor star-galaxy classifier
is unreliable, we assume optical counterparts are extended.} 
\tablenotetext{b}{$z<1$ galaxies with AGES spectra that were classified as
compact in the NDWFS imaging have been reclassified as extended objects.}
\end{deluxetable}

\begin{deluxetable}{ccccc}
\tablecolumns{5}
\tablecaption{Luminosity Function Parameters\label{table:quasarlf}}
\tablehead{ \colhead{Parameter}   &
            \colhead{All}   &
            \colhead{$R-[24]>4.4$\tablenotemark{a}} &
            \colhead{$R-[24]<4.4$\tablenotemark{a}} &
            \colhead{2QZ\tablenotemark{b}}
}
\startdata
$\phi_8(-26, 0) [10^{-7} {\rm Mpc}^{-3}{\rm mag}^{-1}]$  & $4.84  \pm^{5.15}_{3.37}$  & $2.35\pm0.28$   & $2.44\pm0.26$      & -           \\
$\phi_8(-29, 2) [10^{-7} {\rm Mpc}^{-3}{\rm mag}^{-1}]$  & $4.46  \pm 0.53$           & $2.16\pm0.26$   & $2.26\pm0.25$      & -           \\
$\alpha$                                                 & $-2.75 \pm^{0.17}_{0.11}$  & $-2.82\pm0.14$  & $-2.66\pm0.13$     & $-3.31$     \\
$k_1$                                                    & $+1.15 \pm 0.34$           & $+1.15$         & $+1.15$            & $+1.39$     \\
$k_2$                                                    & $-0.34 \pm 0.15$           & $-0.34$         & $-0.34$            & $-0.29$     \\
$k_3$                                                    & $+0.03 \pm 0.02$           & $+0.03$         & $+0.03$            & -           \\
$z_{peak}$                                               & $2.56  \pm 0.27$           & $2.56$          & $2.56$             & 2.4         \\
$z$ range                                                & $1<z<5$                    & $1<z<5$         & $1<z<5$            & $0.4<z<2.1$ \\
Spectroscopic sample size                                & 183                        & 81              & 102                & 15,830      \\
\enddata
\tablenotetext{a}{Evolution parameters have been fixed for the red and blue subsamples.}
\tablenotetext{b}{2QZ $z_{peak}$ value is an extrapolation of the evolution of the $0.4<z<2.1$ luminosity function.}
\end{deluxetable}

\begin{deluxetable}{ccc}
\tablecolumns{3}
\tablecaption{Binned $1/V_{max}$ $z=2$ Quasar Luminosity Function\label{table:binned}}
\tablehead{\colhead{$8~\micron$ absolute}   &
           \multicolumn{2}{c}{Space Density at $z=2$ $({\rm Mpc}^{-3}{\rm mag}^{-1})$\tablenotemark{a}} \\
           \colhead{magnitude range}       &
	   \colhead{$1.0<z<5.0$ Quasars} &
	   \colhead{$1.5<z<2.5$ Quasars}
} 
\startdata
$-27.4<M_8<-27.0$   & $4.3\pm 1.0 \times 10^{-6}$ & -                          \\
$-27.8<M_8<-27.4$   & $2.4\pm 0.5 \times 10^{-6}$ & -                          \\
$-28.2<M_8<-27.8$   & $3.0\pm 0.4 \times 10^{-6}$ & $2.4\pm 0.6 \times 10^{-6}$ \\
$-28.6<M_8<-28.2$   & $1.2\pm 0.2 \times 10^{-6}$ & $1.1\pm 0.2 \times 10^{-6}$ \\
$-29.0<M_8<-28.6$   & $6.1\pm 0.9 \times 10^{-7}$ & $4.5\pm 1.0 \times 10^{-7}$ \\
$-29.4<M_8<-29.0$   & $2.6\pm 0.5 \times 10^{-7}$ & $2.7\pm 0.7 \times 10^{-7}$ \\
$-29.8<M_8<-29.4$   & $2.4\pm 0.5 \times 10^{-7}$ & $2.0\pm 0.6 \times 10^{-7}$ \\
$-30.2<M_8<-29.8$   & $9.2\pm 2.4 \times 10^{-8}$ & $6.8\pm 2.5 \times 10^{-7}$ \\
$-30.6<M_8<-30.2$   & $4.7\pm 1.5 \times 10^{-8}$ & $7.4\pm 2.8 \times 10^{-8}$ \\
$-31.0<M_8<-30.6$   & $<7.2       \times 10^{-8}$ & $<2.8       \times 10^{-7}$ \\
$-31.4<M_8<-31.0$   & $2.3\pm 0.9 \times 10^{-8}$ & $4.9\pm 2.1 \times 10^{-8}$ \\
\enddata
\tablenotetext{a}{Upper limits are $3\sigma$.}
\end{deluxetable}

\end{document}